\PassOptionsToPackage{colorlinks=true,urlcolor=blue,anchorcolor=blue,citecolor=blue,filecolor=blue,linkcolor=blue,menucolor=blue,pagecolor=blue,linktocpage=true,pdfproducer=medialab,pdfa=true,colorlinks,citecolor=tl,linkcolor=blue,urlcolor=magenta,unicode=true}{hyperref}
\documentclass[a4paper,11pt]{article}
\usepackage{url}
\pdfoutput=1
\usepackage{jcappub}
\usepackage[T1]{fontenc}
\usepackage[normalem]{ulem}

\usepackage{graphicx}
\usepackage{xcolor}
\definecolor{tl}{RGB}{0,180,120}
\usepackage[colorlinks,citecolor=tl,linkcolor=blue,urlcolor=magenta,unicode=true]{hyperref}

\newcommand{\added}[1]{\textcolor{black}{#1}}

\title{\added{Modified Expansion and the JWST–CMB Optical Depth Tension in Cosmology inspired by Self-Interaction Gravity}}

\author[a,1]{Subhadeep Mukherjee,\note{Corresponding author.}}
\author[a]{Shashank Shekhar Pandey,}
\author[a]{ and A. S. Majumdar}

\affiliation[a]{S. N. Bose National Centre for Basic Sciences,\\ JD Block, Sector III, Salt Lake, Kolkata-700106, India.}

\emailAdd{shubhadeep.avg@gmail.com}
\emailAdd{shashankpandey7347@gmail.com}
\emailAdd{archan@bose.res.in}

\abstract{\added{Recent JWST observations suggests an earlier and more efficient reionization history, and thus larger Thomson optical depths than inferred from the from the cosmic microwave background (CMB). We test whether a modified expansion history motivated by self-interaction (SI) gravity can reduce this tension while retaining the standard astrophysical description of reionization. The parameters of an SI-motivated depletion function are constrained with a Bayesian MCMC using Union3 Type Ia supernovae, DESI DR2 BAO, and the CMB acoustic scale $(\theta_*)$. We then compute the ionization history and optical depth, including cosmology-dependent remapping of UV luminosity functions. The predicted optical depth drops from $\tau_{\rm CMB}=0.0745\pm0.0001$ in $\Lambda$CDM to $\tau_{\rm CMB} = 0.0643 \pm 0.001$ in the SI-motivated background. This shifts agreement with Planck PR3 to $\sim1.4\sigma$, versus $\sim3\sigma$ in $\Lambda$CDM, easing the tension. While the SI-motivated model slightly improves the best fit to the combined data, information criteria show no significant preference over $\Lambda$CDM.}}

\begin{document}
\maketitle
\flushbottom

\section{Introduction}
\label{Sec:intro}

\par The epoch of reionization marks a key phase in cosmic history, during which ultraviolet radiation from the first stars and galaxies ionized the neutral hydrogen in the intergalactic medium (IGM) \cite{BARKANA2001125, Loeb2013}. It directly affects observations of the cosmic microwave background (CMB) through the Thomson scattering optical depth, $\tau_{\rm{CMB}}$, which quantifies the integrated opacity due to free electrons along the line of sight, making it a sensitive probe of both astrophysical processes and the cosmic expansion history.

\par Recent observations from the \emph{James Webb Space Telescope} (JWST) have significantly revised our understanding of early galaxy formation and the reionization history of the universe. In particular, the unexpectedly high abundance of galaxies at $z\gtrsim9$~\cite{DONNAN2024,Harikane2024}, together with revised ultraviolet luminosity functions (UVLFs) exhibiting higher galaxy number densities and increased ionizing-photon production efficiencies
($\xi_{\rm ion}$)~\cite{Simmonds2023,Atek2024}, reveal a substantially
larger ionizing-photon budget than previously inferred from \emph{Hubble Space Telescope} (HST) observations~\cite{Robertson_2015,Finkelstein_2019}.

\par These developments suggest an earlier and potentially more efficient reionization of the intergalactic medium. Since the Thomson scattering optical depth measured from CMB observations is determined by the integrated free-electron density along the line of sight, earlier ionization generally yields a larger value of $\tau_{\rm{CMB}}$. Consequently, reionization histories inferred from JWST observations may predict optical depths that are higher than those inferred from the \textit{Planck} CMB data, thereby highlighting a potential tension between the two probes \cite{Munoz2024}.

\par Several recent studies have explored possible resolutions to this tension, broadly falling into two complementary directions. On the astrophysical side, modifications to reionization modeling, such as revisions to ultraviolet luminosity functions, ionizing-photon production efficiencies, escape fractions, and the contribution of faint galaxies, have been extensively investigated \cite{Mirocha2023, Mason2023, fererra2023}. However, current analyses indicate that astrophysical adjustments alone may be insufficient. In particular, Mu$\tilde{n}$oz et al.~\cite{Munoz2024} demonstrate that the galaxy populations inferred from JWST observations not only drive cosmic reionization but would also complete it too early, leading to an overproduction of ionizing photons and thereby exacerbating the disagreement with the CMB optical-depth measurements and the Ly$\alpha$ forest constraints. Similar conclusions have also been discussed in the context of post-JWST reionization modeling by \cite{Mirocha2023, Mason2023}.

\par On the cosmological side, an alternative possibility is that the tension originates from the cosmological framework used to interpret both reionization and CMB observations. Several studies have therefore investigated extensions or modifications of the standard cosmological model, such as in \cite{Melia2024}. It has been shown \cite{Jhaveri2025}  that relaxing the CMB lensing constraint can lead to a higher preferred value of the optical depth, $\tau_{\rm{CMB}}$. Similarly, \cite{Giare2023} found that determinations of $\tau_{\rm{CMB}}$ independent of large-scale CMB polarization generally favour larger optical-depth values. Modifications to the primordial power spectrum have also been investigated \cite{Huang2025,Upadhyay_2026}, although such change alone appears insufficient to substantially increase the inferred value of $\tau_{\rm{CMB}}$.

\par While the studies discussed above explore several possible resolutions to the optical-depth tension, most are formulated within the framework of the standard $\Lambda$CDM cosmology.  Despite its remarkable success in describing a broad range of cosmological observations, the $\Lambda$CDM model relies on the assumption that the universe is statistically homogeneous and isotropic on sufficiently large scales. This assumption, however, has increasingly been questioned by both theoretical considerations and observational evidence \cite{Aluri_2023}. Moreover, several observational studies have reported large-scale structures extending over scales significantly larger than those expected within the standard cosmological framework. Analyses of galaxy distributions from the Two-degree Field Galaxy Redshift Survey (2dFGRS) revealed large-amplitude density fluctuations extending to scales of nearly $100\,h^{-1}\,\mathrm{Mpc}$ \cite{{Labini_2009}}. Likewise, investigations based on the Sloan Digital Sky Survey (SDSS) identified statistically significant deviations from $\Lambda$CDM mock catalogues on scales approaching $500\,h^{-1}\,\mathrm{Mpc}$ \cite{wiegand_scale}. The discovery of giant arc-like galaxy structures spanning nearly $1\,\mathrm{Gpc}$ \cite{lopez} has reignited the debate regarding the scale of cosmic homogeneity \cite{Lopez2025, Swala2025}. Such observations have motivated investigations into whether cosmic inhomogeneities and large-scale structure can influence the inferred cosmic expansion history and affect various astrophysical and cosmological observables \cite{pandey_effect_2022, pandey_analyzing_2024, mukherjee_constraining_2025, pandey_viscous_2023, halder_future_2023}. The Buchert backreaction formalism \cite{Buchert2001, Rasanen2006, Wiltshire_2007, Weigand2010, buchertrasanen2012, Buchert2020} leads to
predictions that  align well with recent observational developments \cite{DESI_DR2, Lodha2025}, and also   yields an estimate of the optical depth to reionization that is more consistent with observations than the value obtained in the standard $\Lambda$CDM model.

\par Furthermore, persistent cosmological tensions \cite{Riess_2022,plank_pr3, DIVALENTINO2025}, particularly the Hubble tension \cite{RoyChoudhury2026}, have stimulated renewed interest in cosmological models beyond the standard paradigm \cite{Valentino_2021}. Recent DESI BAO observations \cite{DESI_DR2, Lodha2025} provide evidence favoring a dynamically evolving dark-energy component rather than a strictly constant cosmological constant. Although these indications remain under active investigation \cite{Linder2025}, they motivate the exploration of alternative mechanisms capable of generating the observed late-time acceleration through gravitational dynamics themselves rather than by introducing an additional dark-energy field \cite{Buchert2008, Buchert2020}.

\par Driven by the same consideration that loosening the assumption of exact homogeneity can modify the cosmic expansion history, the self-interaction (SI) gravity framework was introduced in Refs.~\cite{Deur2009, Deur2019}. Building on the observed inhomogeneous and anisotropic distribution of matter in the universe, this approach suggests that the intrinsic nonlinearity of general relativity inevitably leads to gravitational self-interaction effects once large-scale structures have emerged and that these effects can become dynamically relevant on cosmological scales. By analogy with other self-interaction field theories, such as Quantum Chromodynamics (QCD), these self-interaction effects can lead to effective trapping of the field within the bound structure. This mechanism addresses both the dark matter and dark energy phenomena simultaneously. On galactic and cluster scales, this manifests as an enhancement of the effective gravitational attraction, reproducing phenomena commonly attributed to dark matter \cite{Deur2009, Deur2014}. Conversely, the corresponding depletion of the gravitational field outside these structures weakens gravity on large scales, leading to an effective accelerated expansion without invoking a separate dark energy component \cite{Deur2019}. Incorporating self-interaction effects modifies the Friedmann equation and consequently alters the Hubble expansion history.

\par \added{The reionization history depends on the cosmic expansion rate through the relation between redshift and cosmic time and through the calculation of the Thomson optical depth. Hence, a modification of the Hubble parameter $H(z)$ can affect the evolution of the ionized fraction, the timing of reionization, and the resulting optical depth. This dependence is a general cosmological effect and may not not specific to any particular gravitational framework. Different expansion histories can, in principle, produce different reionization histories even when the astrophysical prescriptions are held fixed. SI gravity provides one physically motivated framework in which it could be explored whether a modified cosmological background can affect the apparent discrepancy between the optical depth inferred from JWST-informed reionization histories and that measured from cosmic microwave background (CMB) observations.}

\par \added{In this work, we investigate the consequences of an SI gravity-motivated expansion history for cosmic reionization and the associated Thomson optical depth. Rather than modifying the astrophysical ingredients of reionization, we retain the standard description of ionizing-photon production and galaxy evolution and examine their consequences in the modified background. We construct the reionization history using observationally motivated ultraviolet luminosity functions \cite{DONNAN2024}, together with prescriptions for the escape fraction of ionizing photons \cite{Chisholm2022} and the ionizing-photon production efficiency \cite{Simmonds2023}. The evolution of the ionized hydrogen fraction is computed by solving the standard reionization equation \cite{Madau1999}, with the expansion history entering the conversion between redshift and cosmic time. From the resulting ionization history, we calculate the Thomson scattering optical depth and compare it with CMB constraints. Since the observed ultraviolet luminosity functions are inferred under an assumed reference cosmology, their cosmological dependence is also considered when applying them to a modified expansion history.}

\par \added{The parameters of the adopted SI gravity background prescription are constrained using a joint analysis of the Union3 Type Ia supernova compilation \cite{union3} and the DESI DR2 baryon acoustic oscillation (BAO) measurements \cite{DESI_DR2}, along with a CMB likelihood for the angular acoustic scale($\theta_{*}$). The observationally constrained expansion history is then used to investigate the resulting reionization history and optical depth. We assess whether the modified expansion history can reduce the discrepancy between the optical depth inferred from JWST-informed reionization models \cite{Munoz2024} and the value measured by Planck \cite{plank_pr3}. Our analysis demonstrates that modifications to the cosmological expansion history reduce the discrepancy between the optical depth predicted from JWST-informed reionization models \cite{Munoz2024} and the Planck-measured values \cite{plank_pr3}. The two values agree at the $1.4\sigma$ level. This assessment is performed by considering the consistency between the background evolution and the CMB acoustic scale, and by reviewing the cosmological assumptions underlying the ultraviolet luminosity functions. We also examine whether the resulting ionization history is compatible with independent observational constraints
\cite{ly1,ly2,ly3,ly4,ly5,ly6,ly7,ly8}. The purpose is to determine whether the optical-depth discrepancy can be alleviated within the adopted background prescription, rather than to attribute such an
effect uniquely to SI gravity.}

\par \added{The remainder of this paper is organized as follows. In
Section~\ref{Sec:reionization}, we outline the reionization model,
including the ultraviolet luminosity functions, the ionizing-photon
production efficiency, the escape-fraction prescription, and the
calculation of the Thomson optical depth.
Section~\ref{Sec:si_gravity} introduces the SI gravity motivated
background prescription and describes the modified cosmological
expansion history. In Section~\ref{Sec:par_explrn}, we investigate
the sensitivity of the reionization observables to the parameters
of this prescription, accounting for the cosmological assumptions
underlying the ultraviolet luminosity functions. Section~\ref{Sec:observational_constraints} presents the observational constraints obtained from a joint analysis of the Union3 Type Ia supernova compilation, DESI DR2 BAO measurements, and CMB acoustic scale($\theta_*$), together with the corresponding model comparison between $\Lambda$CDM and SI gravity. In Section~\ref{Sec:jwst_cmb_reconsiln}, we examine the implications of the constrained expansion history for cosmic reionization and compare the resulting ionization history and Thomson optical depth with current observations. Finally,
Section~\ref{Sec:conclusion} summarizes our findings, their limitations, and their implications for further tests of modified cosmological expansion histories.}

\section{Reionization Model}
\label{Sec:reionization}

To study the reionization history, we follow the standard volume-averaged approach and describe the evolution of the ionized hydrogen fraction, $x_{\mathrm{HII}}$ or equivalently, the neutral hydrogen fraction, $x_{\mathrm{HI}} = 1 -x_{\mathrm{HII}}$. The ionization balance equation is given by \cite{Madau1999}

\begin{equation}
\frac{d x_{\mathrm{HII}}}{dt}
=
\frac{\dot{n}_{\mathrm{ion}}}{n_{\mathrm H}}
-
\frac{x_{\mathrm{HII}}}{t_{\mathrm{rec}}},
\label{eq:ionization_balance}
\end{equation}

where the first and second terms on the right-hand side represent the source and sink of ionizing photons, respectively. The competition between these two terms determines the reionization history of hydrogen. The first term corresponds to the ratio of the ionizing photon production rate, $\dot{n}_{ion}$ to the \added{proper} hydrogen number density, $n_H=n_{H0}\times(1+z)^3$ \added{where $n_{H0}$ is the comoving hydrogen number density} given by,
\begin{equation}
n_{\mathrm H0}
=
\frac{\Omega_b \rho_c (1-Y_{\mathrm{He}})}
{m_{\mathrm H}},
\label{eq:hydrogen_density}
\end{equation}
where $\Omega_b$ denotes the \added{present} baryonic matter density parameter, $\rho_c$ represents the current critical density of the universe, $Y_{\rm He}$ is the helium mass fraction, and $m_{\rm H}$ is the mass of the proton.

The second (sink) term quantifies the average recombination rate of hydrogen atoms and is characterized by the recombination timescale, $t_{rec}$ \cite{Shull2012}

\begin{equation}
t_{\mathrm{rec}}
=
\left[
C\,\alpha_B\,(1+\chi_{\mathrm{He}})\,
n_{\mathrm H}
\right]^{-1},
\label{eq:recombination_time}
\end{equation}
where $C$ is the clumping factor and $\alpha_B$ is the case-B recombination coefficient. 

\begin{equation}
\chi_{\mathrm{He}}
=
\frac{n_{\mathrm{He}}}{n_{\mathrm H}}
\simeq
\frac{Y_{\mathrm{He}}}
{4(1-Y_{\mathrm{He}})},
\label{eq:helium_fraction}
\end{equation}
denotes the helium abundance relative to hydrogen. Following \cite{Robertson_2015,Munoz2024,Melia2024}, we adopt a constant clumping factor $C=3$, and the case-B recombination coefficient is evaluated at $T = 2 \times 10^4$ Kelvin.

\par The ionizing photon production rate, $\dot{n}_{\mathrm{ion}}$, appearing in the source term represents the \added{proper} production rate of ionizing photons by star-forming galaxies and is written as,

\begin{equation}
\dot{n}_{\mathrm{ion}}
=
\int
dM_{\rm UV}\,
\Phi_{\rm UV}(M_{\rm UV},z)\,
(1+z)^3\,
\dot{N}_{\rm ion}(M_{\rm UV},z)\,
f_{\rm esc}(M_{\rm UV},z),
\label{eq:ionising_emissivity}
\end{equation}
where every quantity inside the integral is a function of the ultraviolet absolute magnitude $M_{\rm UV}$. In this expression, $\Phi(M_{\rm UV},z)$ denotes the galaxy ultraviolet luminosity function, $L_{\rm UV}(M_{\rm UV})$ is the rest-frame ultraviolet luminosity associated with $M_{\rm UV}$, $\xi_{\rm ion}(M_{\rm UV},z)$ represents the ionizing photon production efficiency, and $f_{\rm esc}(M_{\rm UV},z)$ denotes the fraction of hydrogen-ionizing photons produced by a galaxy that escape from the galaxy into the IGM. The integration is performed down to a cutoff ultraviolet magnitude, $M_{\rm UV}^{\rm cutoff}$, which represents the faintest galaxies assumed to contribute to the ionizing-photon budget and is, in general, a free parameter of the reionization model. Current JWST observations directly probe galaxies down to approximately $M_{\rm UV}\simeq-15$ \cite{Atek2018}, while theoretical studies predict that the ultraviolet luminosity function should eventually turn over at fainter magnitudes owing to radiative and supernova feedback in low-mass galaxies \cite{Shapiro2004}. However, the location of this turnover has not yet been observationally determined. The study \cite{Munoz2024} treated this limiting magnitude as a free parameter in order to investigate its impact on the reionization history. In the present work, however, we adopt the conservative value $M_{\rm UV}^{\rm cutoff}=-14.6$, following the choice of \cite{Melia2024}, who argued that this value is consistent with the current observational limits while avoiding an extrapolation to substantially fainter galaxies. 

 \added{Throughout the reionization calculation, the ultraviolet luminosity function $\Phi_{\rm UV}(M_{\rm UV},z)$ is defined as a comoving number density per unit absolute magnitude. Accordingly, the factor $(1+z)^3$ in \autoref{eq:ionising_emissivity} converts the comoving galaxy number density into a proper number density before evaluating the ionizing-photon emissivity. The hydrogen number density $n_{\rm H}$ and the electron number density $n_e$ entering the recombination and Thomson optical-depth calculations are likewise proper number densities. With these conventions, the source and recombination terms in the ionization balance equation are expressed consistently per hydrogen atom in the physical intergalactic medium.} In this work we adopt the UVLF calibration of Bouwens et al.~\cite{Bouwens2021} for $z\le9$, while for $z>9$ we use the JWST-based determination of Donnan et al.~\cite{DONNAN2024}.

The ionizing photon production rate of an individual galaxy is expressed as,

\begin{equation}
\dot{N}_{\rm ion}
=
L_{\rm UV}\,
\xi_{\rm ion},
\label{eq:ionising_photons}
\end{equation}
where $L_{\rm UV}$ is the rest-frame UV luminosity inferred from the galaxy absolute magnitude $M_{\rm UV}$ following the AB magnitude system \cite{OkeGunn1983}, and $\xi_{\rm ion}$ is the ionizing photon production efficiency.
For pre-JWST analyses, the ionizing efficiency is commonly assumed to follow $\log_{10}\left(\frac{\xi_{\rm ion}}{{\rm Hz\,erg^{-1}}}\right)=25.2$ \cite{Robertson2013}. Recent JWST observations suggest significantly enhanced ionizing photon production efficiencies, particularly in faint, high-redshift galaxies. Following \cite{Simmonds2023,Atek2024,Munoz2024}, we adopt

\begin{equation}
\log_{10}
\left(
\frac{\xi_{\rm ion}}
{{\rm Hz\,erg^{-1}}}
\right)
=
25.8
+
0.11\left(M_{\rm UV}+17\right)
+
0.05\left(z-7\right),
\label{eq:xiion_jwst}
\end{equation}
where the relation is capped at $z=9$ and $M_{\rm UV}=-16.5$ to avoid extrapolation beyond the observationally constrained range.

The final ingredient entering the ionizing emissivity is the escape fraction, $f_{\rm esc}$, which denotes the fraction of hydrogen-ionizing photons that escape from galaxies into the intergalactic medium and contribute to cosmic reionization. The escape fraction remains one of the least constrained quantities at high redshift because ionizing photons are readily absorbed by the neutral interstellar medium. Nevertheless, observations of low-redshift galaxies that resemble those expected during the epoch of reionization reveal a strong correlation between the escape fraction and the ultraviolet continuum slope, $\beta_{\rm UV}$ \cite{Flury2022, Begley2022, Chisholm2022}. Following the prescription of \cite{Chisholm2022}, we write

\begin{equation}
f_{\rm esc}
=
A_f
\,10^{\,b_f\,\beta_{\rm UV}},
\label{eq:fesc}
\end{equation}
where, $A_f = 1.3\times10^{-4},\ b_f=-1.22$. The ultraviolet continuum slope, $\beta_{\rm UV}$, is inferred from the observed $\beta_{\rm UV}$--$M_{\rm UV}$ relation \cite{Zhao2024}. To avoid extrapolation beyond the calibrated range, we cap $\beta_{\rm UV}$ at $-2.7$.

Once the ionized hydrogen fraction has been obtained by solving
\autoref{eq:ionization_balance}, the corresponding CMB Thomson optical depth can be calculated as \cite{Munoz2024},

\begin{equation}
\tau_{\rm CMB}
=
\int
n_e(z)\,
\sigma_T\,
dR,
\label{eq:tau}
\end{equation}
where
$R$
is the proper distance and
$\sigma_T$
is the Thomson scattering cross section.
The proper electron number density is given by,

\begin{equation}
n_e
=
\left(
1+
\frac{\alpha\,Y_{\rm He}}
{4(1-Y_{\rm He})}
\right)
x_{\rm HII}
n_{\rm H}, \hspace{0.5 cm} \text{where} \hspace{0.5 cm} \alpha=
\begin{cases}
1, & z>4,\\
2, & z\le4,
\end{cases}
\label{eq:electron_density}
\end{equation}
accounts for the ionization state of helium. Following \cite{Melia2024, Munoz2024}, we assume that HeI reionization tracks hydrogen reionization, while HeII reionization occurs instantaneously at $z=4$.

Equations \autoref{eq:ionization_balance} - \autoref{eq:electron_density}
fully specify the reionization model adopted in this work. In the numerical implementation, the ionization balance equation is solved after transforming the time derivative into a redshift derivative (see \autoref{Sec:par_explrn}). We assume that the intergalactic medium is initially neutral at the starting redshift of the calculation, and the resulting ionization history is subsequently used to evaluate the CMB Thomson optical depth.

\section{Self-Interaction Gravity}
\label{Sec:si_gravity}

General relativity is intrinsically non-linear. In the field-theoretic formulation of gravity, special relativity requires the Newtonian scalar mass density to be replaced by the stress-energy tensor. Hence, a long-range spin-2 gravitational field must couple to the full stress-energy tensor rather than to a scalar source alone. Consistency, therefore, requires the gravitational field to couple to its own stress-energy tensor in addition to that of matter, giving rise to gravitational self-interaction \cite{Deser1970, Deser2010}.

Iterating this self-coupling to all orders reproduces the full non-linear Einstein field equations. Motivated by the analogous role of self-interaction in Quantum Chromodynamics (QCD), where gluon self-coupling generates non-perturbative phenomena such as confinement, it has been proposed that the self-interaction term already present in general relativity may have observable consequences on astrophysical and cosmological scales \cite{Deur2009, Deur2019}.

In particular, gravitational self-interactions have been shown to enhance the effective binding within sufficiently massive and anisotropic systems, providing an alternative explanation for the missing mass phenomenon in galaxies and galaxy clusters without invoking non-baryonic dark matter \cite{Deur2009, Deur2014}. Extending this idea to cosmology, the progressive trapping of gravitational fields during structure formation was argued to weaken the effective gravitational field on scales larger than the bound structures, thereby modifying the background expansion history and reproducing the Type Ia supernova observations commonly interpreted as evidence for dark energy \cite{Deur2019}.

The central concept of the model is the trapping of a gravitational field. As matter collapses into galaxies, groups, and clusters, a growing fraction of the gravitational field becomes effectively confined within these structures. The enhanced binding within collapsed systems is accompanied by a corresponding depletion of the gravitational field on large scales. As a result, gravity becomes effectively stronger inside bound structures while its large-scale cosmological influence is progressively reduced as structure formation proceeds.

In the standard derivation of the Friedmann equations, the universe is assumed to be perfectly homogeneous and isotropic. Under these assumptions, the effects of field trapping are absent. Relaxing these assumptions introduces additional anisotropy and inhomogeneity terms into the Einstein equations. These contributions can be absorbed into an effective depletion factor, $D(z)$, yielding the modified Friedmann equation

\begin{equation}
\dot{a}^{\,2}+k
=
\frac{8\pi G}{3}\,
D(z)\,
\rho\,a^{2},
\label{eq:modified_friedmann}
\end{equation}
where $a$ is the scale factor, $G$ is the Newton gravitational constant, $k$ denotes the spatial curvature and $\rho$ is the total energy density of the universe. In general, it contains contributions from non-relativistic matter, radiation, and the cosmological constant, the latter being treated as an effective fluid with constant energy density.

Since anisotropies generated by structure formation may affect relativistic matter, non-relativistic matter, and the cosmological constant differently, the depletion factors and energy densities are generalized as vectors, $ D \rightarrow (D_{\mathrm{M}},D_{\mathrm{R}},D_{\Lambda}),
\ \rho \rightarrow
(\rho_{\mathrm{M}},\rho_{\mathrm{R}},\rho_{\Lambda}),
$ where the subscripts M, R, and $\Lambda$ denote matter, radiation, and the cosmological constant, respectively \cite{Deur2019}. The quantity $D(z)\rho$ appearing in Eq.~\eqref{eq:modified_friedmann} is then interpreted as the scalar product of these vectors,

\begin{equation}
D\cdot\rho
=
D_{\mathrm{M}}\rho_{\mathrm{M}}
+
D_{\mathrm{R}}\rho_{\mathrm{R}}
+
D_{\Lambda}\rho_{\Lambda},
\label{eq:scalar_product}
\end{equation}
and the modified Friedmann equation becomes

\begin{equation}
\dot{a}^{\,2}+k
=
\frac{8\pi G}{3}
\left(
D_{\mathrm{M}}\rho_{\mathrm{M}}
+
D_{\mathrm{R}}\rho_{\mathrm{R}}
+
D_{\Lambda}\rho_{\Lambda}
\right)a^{2}.
\label{eq:modified_friedmann_components}
\end{equation}

Following \cite{Deur2019}, it is convenient to introduce the screened density fractions,

\begin{equation}
\Omega_i^{*}(z)=\Omega_i\,D_i(z),
\label{eq:screened_density}
\end{equation}
where, $\Omega_i=\frac{8\pi G\,\rho_{i,0}}{3H_0^2}$ are the conventional density parameters, with $H_0$ being the present-day Hubble parameter and $\rho_{i,0}$ representing the present-day energy density of the corresponding component. The screened density fractions characterize the effective contribution of each component to the cosmic expansion after accounting for gravitational field depletion. They should therefore not be interpreted as the actual matter densities of the universe, but rather as dynamical quantities governing the expansion history.

Defining the dimensionless expansion function

\begin{equation}
E(z)\equiv\frac{H(z)}{H_0},
\label{eq:Ez_definition}
\end{equation}
where $H(z)$ is the Hubble parameter at redshift $z$, the modified Friedmann equation may be written as

\begin{equation}
E^{2}(z)
=
\Omega_{\mathrm{M}}^{*}(z)(1+z)^3
+
\Omega_{\mathrm{R}}^{*}(z)(1+z)^4
+
\Omega_{\Lambda}^{*}(z)
+
\Omega_{k}(1+z)^2,
\label{eq:Ez_general}
\end{equation}
where

\begin{equation}
\Omega_k=-\frac{k}{a_0^2H_0^2},
\label{eq:omega_k}
\end{equation}
is the curvature density parameter, with $a_0$ denoting the present-day scale factor. After the matter--radiation equality epoch ($z_{\rm eq}\simeq3400$) and assuming the cosmological constant is absent, $D_{\Lambda}\simeq0,(\rho_{\Lambda}\simeq0),$ one has $D\simeq(D_{\rm M},0,0),\ \rho\simeq(\rho_{\rm M},0,0)$ \cite{Deur2019}. The dominant depletion therefore arises from the non-relativistic matter component alone, and the expansion history reduces to,
\begin{equation}
E^{2}(z) = \Omega_{\rm M}^{*}(z)(1+z)^3 + \Omega_k(1+z)^2.
\label{eq:Ez_matter}
\end{equation}

Since the SI gravity framework adopts the Einstein–de Sitter universe as the reference FLRW cosmology, the present-day matter density satisfies $\Omega_{\rm M}=1$ with the assumption of  $\Omega_{\Lambda}=0 \ and \ \Omega_{\rm R}\ll1$ at late time (i.e. $z<<3400$). Finally, with $\Omega_{\rm M}^{*}(0)=D_{\rm M}(0)$ because $\Omega_{\rm M} = 1$, together with the normalization condition $E(0)=1$, the curvature density parameter can be estimated as,

\begin{equation}
\Omega_k
=
1-D_{\rm M}(0).
\label{eq:omega_k_final}
\end{equation}
A potential concern is that the SI gravity model predicts a non-zero value of the effective curvature parameter $\Omega_k$, whereas the standard $\Lambda$CDM interpretation of CMB observations is consistent with a spatially flat Universe \cite{plank_pr3}. However, this does not necessarily imply a disagreement with the observations, since the background dynamical evolution in SI gravity differs from that of the standard $\Lambda$CDM model, as also argued in the original SI gravity paper \cite{Deur2019}. Furthermore, it has been shown that the same SI gravity background with a non-zero $\Omega_k$ can reproduce the observed CMB temperature power spectrum competitively with the $\Lambda$CDM model \cite{Deur2022}. \added{Nevertheless, those calculations were performed using Weinberg's hydrodynamic approximation
\cite{Weinberg2008}. A definitive assessment of the consistency of the SI gravity framework with CMB observations therefore requires a treatment based on the full perturbation equations. Such an analysis lies beyond the scope of the present work.}

\par The entire cosmological impact of self-interaction gravity is therefore encoded in the redshift evolution of the screened matter density, or equivalently, the depletion function $D_M(z)$.
Since a first-principles calculation of the depletion function is not currently available, the redshift evolution of $D_M(z)$ is modelled phenomenologically using observational information from the history of structure formation. Following \cite{Deur2022,Sargent2024}, the depletion function is parameterized as,
\begin{equation}
D_M(z)
=
1-
\left(
1+e^{(z-z_g)/\tau}
\right)^{-1}
+
A\,e^{-z/b},
\label{eq:depletion_function}
\end{equation}
where $z_g$ corresponds to the redshift halfway through the galaxy formation epoch, $\tau$ characterizes the duration of this epoch, $A$ quantifies the fraction of structures whose subsequent evolution increases their symmetry, and $b$ determines the corresponding timescale. 

This functional form captures the expected behavior of gravitational field trapping during structure formation. Specifically, $D_M(z)\approx1$ in the early homogeneous and isotropic universe, \added{while} the progressive formation of galaxies, groups, and clusters leads to a reduction in $D_M(z)$ at later times. The exponential term accounts for the partial restoration of symmetry in evolved structures, which weakens field trapping and produces a  increase in $D_M(z)$ toward the present epoch. The self-interaction gravity framework has been proposed as a unified explanation for several cosmological and astrophysical observations. Enhanced gravitational binding within galaxies can reproduce approximately flat rotation curves and cluster dynamics without invoking non-baryonic dark matter \cite{Deur2009, Deur2014}, while the large-scale suppression of gravity can mimic the late-time accelerated expansion conventionally attributed to dark energy.

\added{In the present analysis, the depletion function
$D_M(z)$ is parameterized phenomenologically to construct an expansion history motivated by the SI gravity framework. The purpose is to investigate how this background prescription affects the evolution of the ionized hydrogen fraction and the resulting Thomson optical depth. This parameterization does not constitute a first-principles derivation of the complete background and perturbation equations of SI gravity. Hence, the results obtained below are conditional on the adopted expansion history and should not be interpreted as unique predictions of a fully specified SI gravity cosmology. More generally, the dependence of the reionization history and Thomson optical depth on $H(z)$ is not specific to SI gravity; other modified expansion histories can also affect these observables.}

\par \added{In this work, we investigate how the expansion history associated with the phenomenological depletion function influences cosmic reionization. Since the Hubble parameter $H(z)$ enters the evolution of the ionized hydrogen fraction and the calculation of the Thomson optical depth, variations in the depletion parameters can alter both the ionization history and the resulting optical depth.}

\par \added{To this end, we explore the parameter space of the
depletion function described in
Eq.~\eqref{eq:depletion_function}. The expansion histories are constrained using the Union3 Type Ia supernova observations \cite{union3}, DESI DR2 BAO measurements \cite{DESI_DR2}, and the CMB acoustic scale constraint described in Section~\ref{subsec:cmb_likelihood}. The resulting constraints are used to assess the implications of the allowed expansion histories for the reionization history and cumulative Thomson optical depth.}

\par \added{Finally, we examine whether the resulting optical depth is compatible with CMB observations and whether the corresponding ionization history agrees with independent observational constraints. Any reduction in the optical-depth discrepancy is interpreted largely as a consequence of the adopted expansion history. }


\section{Parameter Exploration} 
\label{Sec:par_explrn}

In \autoref{Sec:reionization}, we have described the reionization formalism governing the evolution of the ionized fraction and the Thomson scattering optical depth, while \autoref{Sec:si_gravity} \added{presented the phenomenological, SI gravity-motivated expansion history. We now combine these two ingredients by incorporating the corresponding Hubble evolution $H(z)$ into the reionization equations.} Since the ionization equation is naturally expressed in terms of cosmic time, it is convenient to rewrite it in terms of redshift as

\begin{equation}
\frac{d}{dt} = -{(1+z)H(z)}\frac{d}{dz},
\label{eq:dt_dz}
\end{equation}

Substituting this relation into \autoref{eq:ionization_balance} introduces an explicit dependence on the modified Hubble parameter $H(z)$, yielding the modified redshift evolution equation for the ionized fraction. In the numerical implementation, this equation is solved from an initial redshift
$z_{\max}=30$ to the present epoch ($z=0$), assuming an initially neutral
intergalactic medium, $x_{\rm HII}(z_{\max})=0$. Likewise, the Thomson scattering optical depth $\tau_{\rm CMB}$ may be expressed as

\begin{equation}
\tau_{\rm CMB}
=
\int_{0}^{z_{\max}}
\frac{n_e(z)\,\sigma_T\,c}
{(1+z)H(z)}
\,dz.
\label{eq:tau_parameter}
\end{equation}

The total CMB optical depth reported throughout this work is obtained by evaluating Eq.~(\ref{eq:tau_parameter}) at $z=z_{\max}$, i.e., $\tau_{\rm CMB}=\tau_{\rm CMB}(z_{\max})$. Thus, both the ionization history and the optical depth depend directly on the cosmological expansion history. Any modification of $H(z)$ arising from the depletion function propagates into the predicted reionization observables. \added{The remaining inputs
are held fixed and identical in the $\Lambda$CDM and SI gravity calculations: the physical baryon density is $\omega_{\rm b}\equiv\Omega_{\rm b}h^2=0.0224$ and the helium mass fraction is $Y_{\rm He}=0.24$. Since $\rho_c\propto h^2$, the hydrogen density $n_{\rm H}$ depends on $\omega_{\rm b}$ and not separately
on $H_0$; the dependence on $H_0$ enters only through $H(z)$ and the distances. The value of $\omega_{\rm b}$ is adopted from Planck  (see Section ~\ref{subsec:cmb_likelihood} for the interpretation of the matter content).} 

\par \added{In addition to its direct influence on the reionization history, the cosmological expansion history also affects the interpretation of the observational ultraviolet luminosity functions (UVLFs) used to calculate the ionizing emissivity. The UVLFs adopted in this work were inferred assuming fiducial cosmologies, and their inferred absolute magnitudes and comoving number densities may therefore change when they are interpreted in the SI gravity-motivated background \cite{Bouwens2021,DONNAN2024}. We describe below an approximate transformation of the published UVLFs that accounts for the changes in luminosity distance and comoving volume.}

\added{The absolute magnitude inferred from an observed flux depends on the luminosity distance through the distance-modulus relation \cite{Hogg_1999}. For the same observed apparent magnitude and assuming identical spectral and \(K\)-correction conventions, the absolute magnitudes inferred under the modified and reference cosmologies are related by}
\added{\begin{equation}
M_{\rm UV}^{\rm SI}
=
M_{\rm UV}^{\rm ref}
-
5\log_{10}\left[
\frac{D_L^{\rm SI}(z)}
     {D_L^{\rm ref}(z)}
\right],
\label{eq:uvlf_magnitude_conversion}
\end{equation}}
\added{where \(D_L^{\rm SI}(z)\) and \(D_L^{\rm ref}(z)\) denote the corresponding luminosity distances. At a fixed redshift, the distance ratio is independent of the galaxy's absolute magnitude and therefore produces an additive shift of the magnitude scale.}

\added{The inferred comoving number density also depends on the cosmological volume assigned to the observed sample. The differential comoving volume element $dV_c$ is given by the standard expression, \cite{Hogg_1999}}
\added{\begin{equation}
\frac{dV_c}{dz\,d\Omega}
=
\frac{c\mathcal{D}_M^2(z)}{H(z)},
\label{eq:uvlf_volume_element}
\end{equation}}
\added{where \(c\) is the speed of light, $\mathcal{D}_M(z)$ is the transverse comoving distance, and $H(z)$ is the Hubble expansion rate. Thus, differences in the distance and expansion histories of the reference $\Lambda$CDM and SI gravity cosmologies lead to different comoving volumes for the same redshift interval and solid angle}

\added{Following the standard definition of the differential luminosity function and its relation to the expected galaxy counts in a comoving survey volume \cite{Sudek_2022}, we require the differential number of galaxies in corresponding magnitude and redshift intervals to remain invariant when the same observed sample is interpreted under the reference and modified cosmologies. This gives}
\added{\begin{equation}
\begin{aligned}
dN
&=
\Phi_{\rm ref}(M_{\rm UV}^{\rm ref},z)\,
dM_{\rm UV}^{\rm ref}\,dV_c^{\rm ref}\\
&=
\Phi_{\rm SI}(M_{\rm UV}^{\rm SI},z)\,
dM_{\rm UV}^{\rm SI}\,dV_c^{\rm SI},
\end{aligned}
\label{eq:uvlf_number_conservation}
\end{equation}}
\added{where $dN$ denotes the differential number of galaxies in the sample, and $\Phi_{\rm ref}$ and $\Phi_{\rm SI}$ are the UVLFs, expressed as the comoving number densities of galaxies per unit absolute-magnitude interval, in the reference $\Lambda$CDM and SI gravity cosmologies, respectively. The quantities $dM_{\rm UV}^{\rm ref}$ and $dM_{\rm UV}^{\rm SI}$ denote the corresponding magnitude intervals, while $dV_c^{\rm ref}$ and $dV_c^{\rm SI}$ are the associated comoving volume elements. Both volume elements refer to the same redshift interval $dz$ and survey solid angle $d\Omega$, and the magnitudes refer to the same observed galaxies interpreted using the respective luminosity distances.}

\added{Rearranging Eq.~\eqref{eq:uvlf_number_conservation} gives}
\added{\begin{equation}
\begin{aligned}
\Phi_{\rm SI}(M_{\rm UV}^{\rm SI},z)
={}&
\Phi_{\rm ref}(M_{\rm UV}^{\rm ref},z)
\left|
\frac{dM_{\rm UV}^{\rm ref}}
     {dM_{\rm UV}^{\rm SI}}
\right|\\
&\times
\frac{dV_c^{\rm ref}}{dV_c^{\rm SI}}.
\end{aligned}
\label{eq:uvlf_general_transformation}
\end{equation}}
\added{The magnitude-bin Jacobian can be obtained by differentiating Eq.~\eqref{eq:uvlf_magnitude_conversion} at fixed redshift. Since the luminosity-distance ratio is independent of absolute magnitude, we obtain}
\added{\begin{equation}
\left|
\frac{dM_{\rm UV}^{\rm ref}}
     {dM_{\rm UV}^{\rm SI}}
\right|=1.
\label{eq:uvlf_magnitude_jacobian}
\end{equation}}
\added{Hence, no additional magnitude-bin Jacobian appears in the transformation. Moreover, because the same \(dz\) and \(d\Omega\) are used in both cosmologies, the ratio of the differential comoving volumes is equal to the ratio of the corresponding volume elements per unit redshift and solid angle. Substituting Eq.~\eqref{eq:uvlf_volume_element} into Eq.~\eqref{eq:uvlf_general_transformation} therefore yields}
\added{\begin{equation}
\begin{aligned}
\Phi_{\rm SI}(M_{\rm UV}^{\rm SI},z)
\simeq{}&
\Phi_{\rm ref}(M_{\rm UV}^{\rm ref},z)\\
&\times
\frac{
\left(dV_c/dz\,d\Omega\right)_{\rm ref}
}{
\left(dV_c/dz\,d\Omega\right)_{\rm SI}
}\\[2pt]
={}&
\Phi_{\rm ref}(M_{\rm UV}^{\rm ref},z)
\left(\frac{\mathcal{D}_M^{\rm ref}(z)}
           {\mathcal{D}_M^{\rm SI}(z)}\right)^2
\frac{H_{\rm SI}(z)}{H_{\rm ref}(z)}.
\end{aligned}
\label{eq:uvlf_cosmology_conversion}
\end{equation}}
\added{Here, $\mathcal{D}_M^{\rm ref}(z)$ and $\mathcal{D}_M^{\rm SI}(z)$ denote the transverse comoving distances in the reference and SI gravity cosmologies, respectively, while $H_{\rm ref}(z)$ and $H_{\rm SI}(z)$ are the corresponding Hubble expansion rates at redshift $z$. The approximate equality in Eq.~\eqref{eq:uvlf_cosmology_conversion} reflects the assumptions involved in applying the count-conservation relation to published UVLF estimates, rather than an additional approximation in the algebraic derivation.}

\added{In particular, this remapping assumes that the same observed galaxy sample, redshift interval, and survey solid angle are retained, that the magnitude transformation is an additive shift at fixed redshift, and that the selection and completeness corrections can be treated as unchanged. The relation is differential in redshift; for a finite redshift bin, the cosmology-dependent corrections must be evaluated across the bin or using an explicitly specified representative redshift. These assumptions matter because published UVLF \cite{Bouwens2021,DONNAN2024} estimates include survey-dependent selection functions, completeness adjustments, and effective volumes. A complete recalibration would require recomputing those observational ingredients in the modified cosmology. Equation~\eqref{eq:uvlf_cosmology_conversion} should therefore be interpreted as an approximate cosmological remapping of the published UVLF parameterizations, not as a full reanalysis of the original galaxy catalogues.}

\added{In applying this remapping to the reionization calculation, the reference UVLF is evaluated at the corresponding reference absolute magnitude, while the magnitude-dependent ionizing-photon production efficiency, \(\xi_{\rm ion}\), and escape fraction, \(f_{\rm esc}\), are evaluated consistently using the adopted modified-cosmology magnitude.}

 \added{Having shown how the modified expansion rate and the cosmology-dependent UVLF prescription feed into the reionization calculation, we now assess how the reionization completion redshift—when the intergalactic medium is fully ionized—and the Thomson optical depth vary with the phenomenological depletion-function parameters, before constraining them with cosmological observations. For this qualitative sensitivity analysis, we adopt a fiducial Hubble constant $H_0 = 70~{\rm km\,s^{-1}\, Mpc^{-1}}$. Since we focus on relative parameter dependence, this particular selection of $H_0$ does not influence the qualitative behavior of the results.}

The adopted parameterization consists of two physically distinct components: the pair $(z_g,\tau)$, which determines the onset and duration of gravitational field depletion during structure formation, and the pair $(A,b)$, which describes the subsequent late-time evolution of the depletion function. Accordingly, we investigate these two parameter pairs separately through contour maps of the reionization completion redshift and the Thomson scattering optical depth. Cross-combinations, such as $(z_g, A)$ or $(\tau, b)$, involving mixed parameters belong to different physical components of the depletion function and are therefore not considered in this qualitative sensitivity analysis. Their statistical correlations are instead explored through the full four-dimensional MCMC analysis presented in the next section (\autoref{Sec:observational_constraints}).

    
    


\begin{figure*}
    \centering
    \begin{tabular}{cc}
	\includegraphics[width=0.5\textwidth]{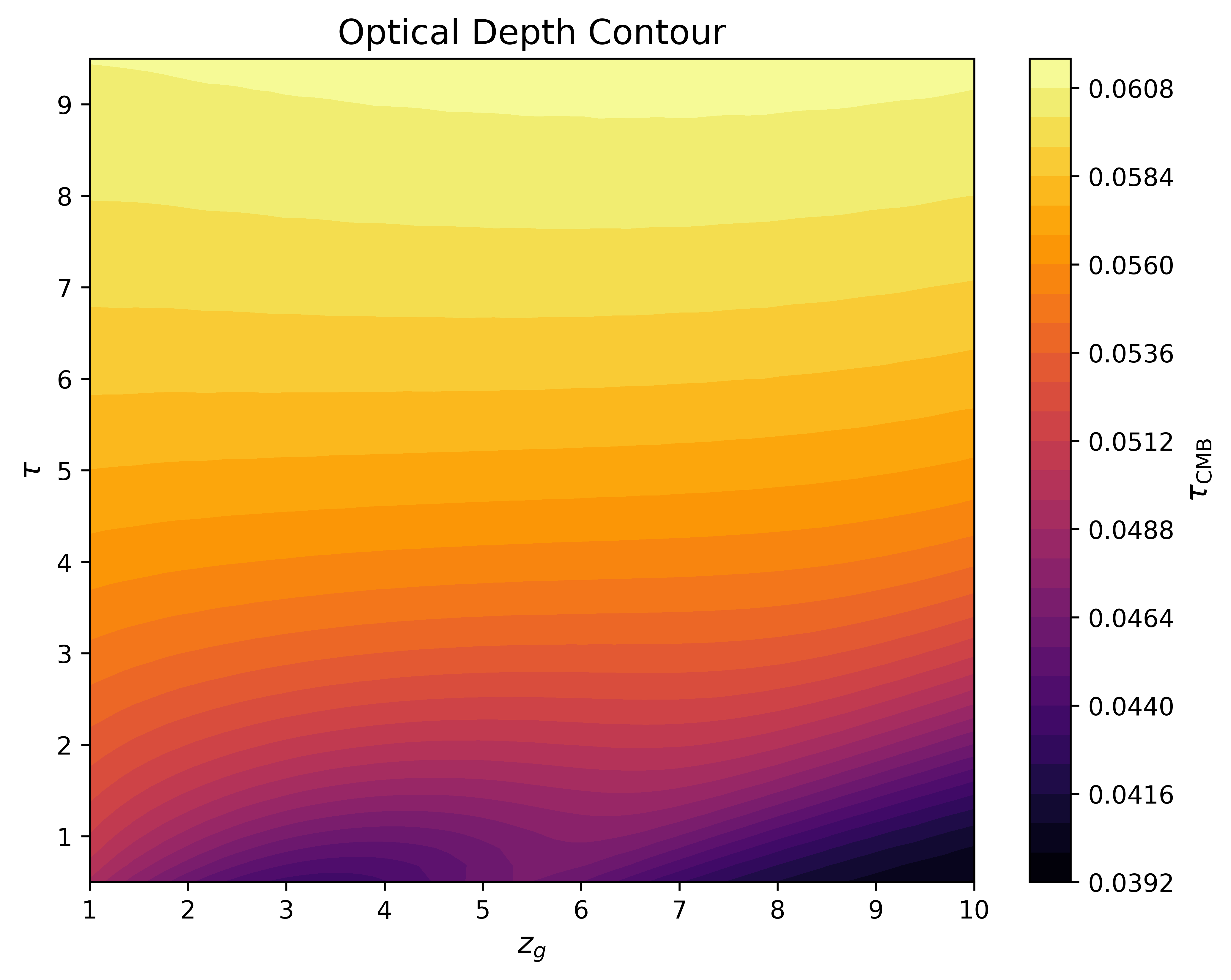}\label{fig:tau_contour}&
	\includegraphics[width=0.5\textwidth]{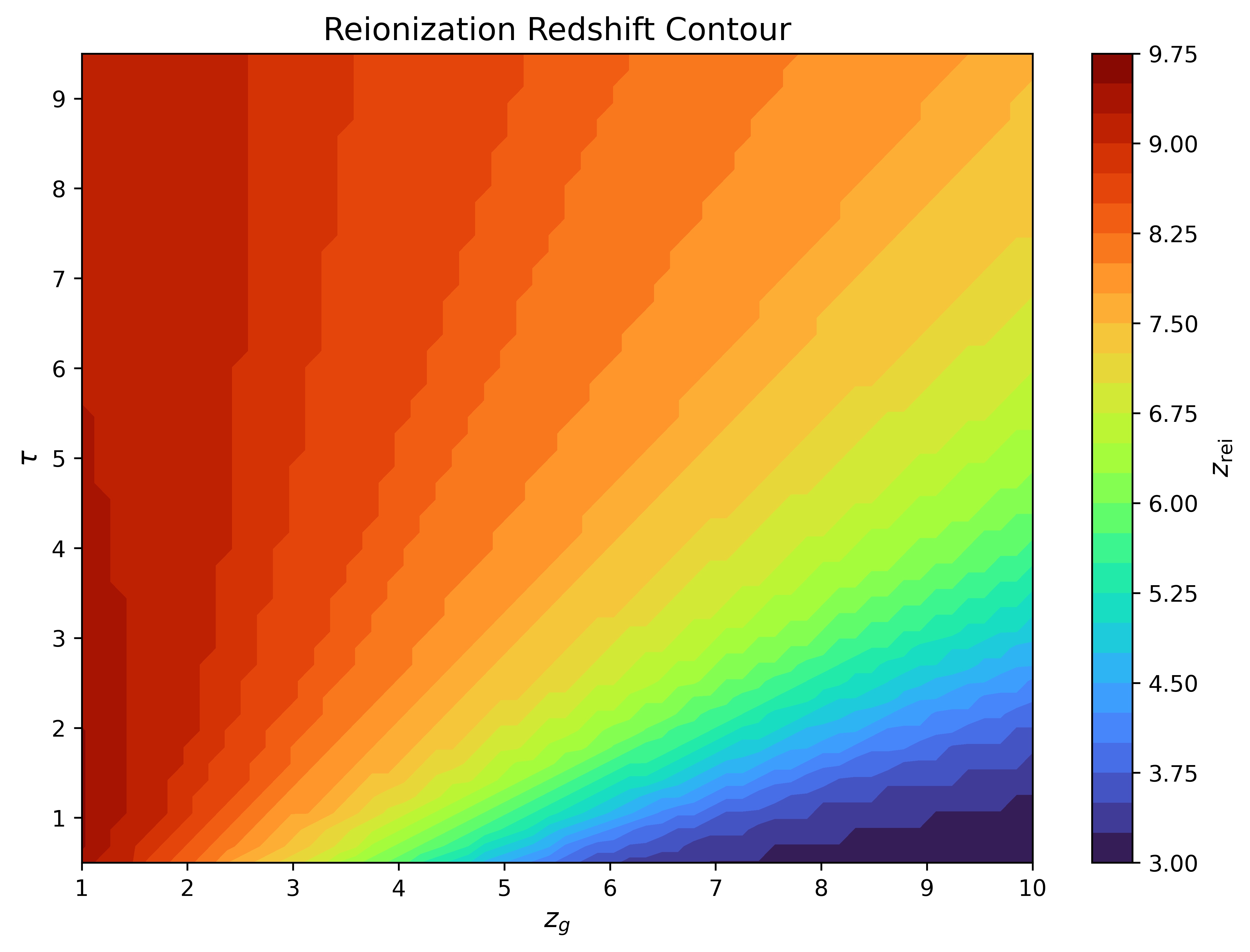}\label{fig:zrei_contour}\\
	(a) Contours of the Thomson optical depth $\tau_{\rm{CMB}}$ & (b) Contours of the reionization redshift $z_{\rm rei}$\\ 
    in the $(z_g,\tau)$ parameter space. & in the $(z_g,\tau)$ parameter space.\\
    \end{tabular}
\caption{\label{fig:zg_tau_contours} Dependence of the reionization observables on the self-interaction gravity parameters $z_g$ and $\tau$, with all other model parameters fixed. The left panel shows the predicted Thomson optical depth, while the right panel shows the corresponding reionization redshift.}
\end{figure*}

\added{\autoref{fig:zg_tau_contours} illustrates the sensitivity of the Thomson optical depth, $\tau_{\rm CMB}$, and the reionization completion redshift, $z_{\rm rei}$, to the depletion-function parameters $(z_g,\tau)$, with the remaining parameters fixed at their fiducial values $A=0.50$ and $b=0.40$. The two contours exhibit notably different patterns of parameter dependence. The optical depth varies predominantly with $\tau$, showing comparatively weaker variation across $z_g$, whereas $z_{\rm rei}$ is sensitive to both parameters. In particular, the dependence of $z_{\rm rei}$ on $z_g$ becomes more pronounced as $z_g$ increases, while its variation is comparatively weak at low $z_g$. The contrast between these contours suggests that a substantial variation in the reionization completion redshift does not necessarily translate into a comparably large variation in the integrated optical depth. A possible explanation is the combined influence of the ionization history and the modified expansion rate. Specifically, the $\tau_{\rm CMB}$ depends on the free-electron density and the expansion rate through the integral \autoref{eq:tau_parameter}. Hence, changes in the ionization history induced by variations in $(z_g,\tau)$ may be partially compensated by the corresponding changes in $H(z)$, reducing the net variation in the integrated optical depth.}

    
    




\begin{figure*}
    \centering
    \begin{tabular}{cc}
	\includegraphics[width=0.5\textwidth]{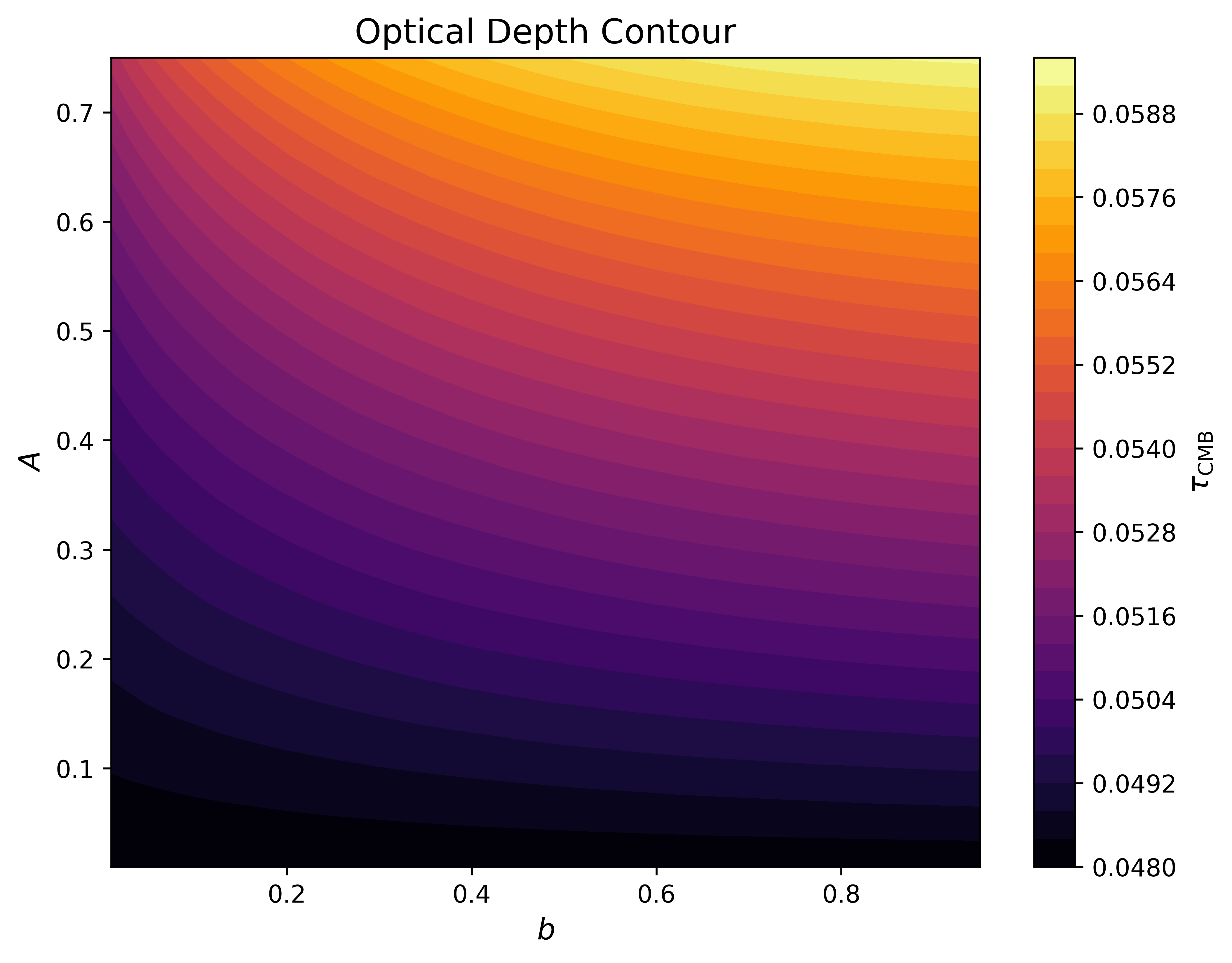}& 
	\includegraphics[width=0.5\textwidth]{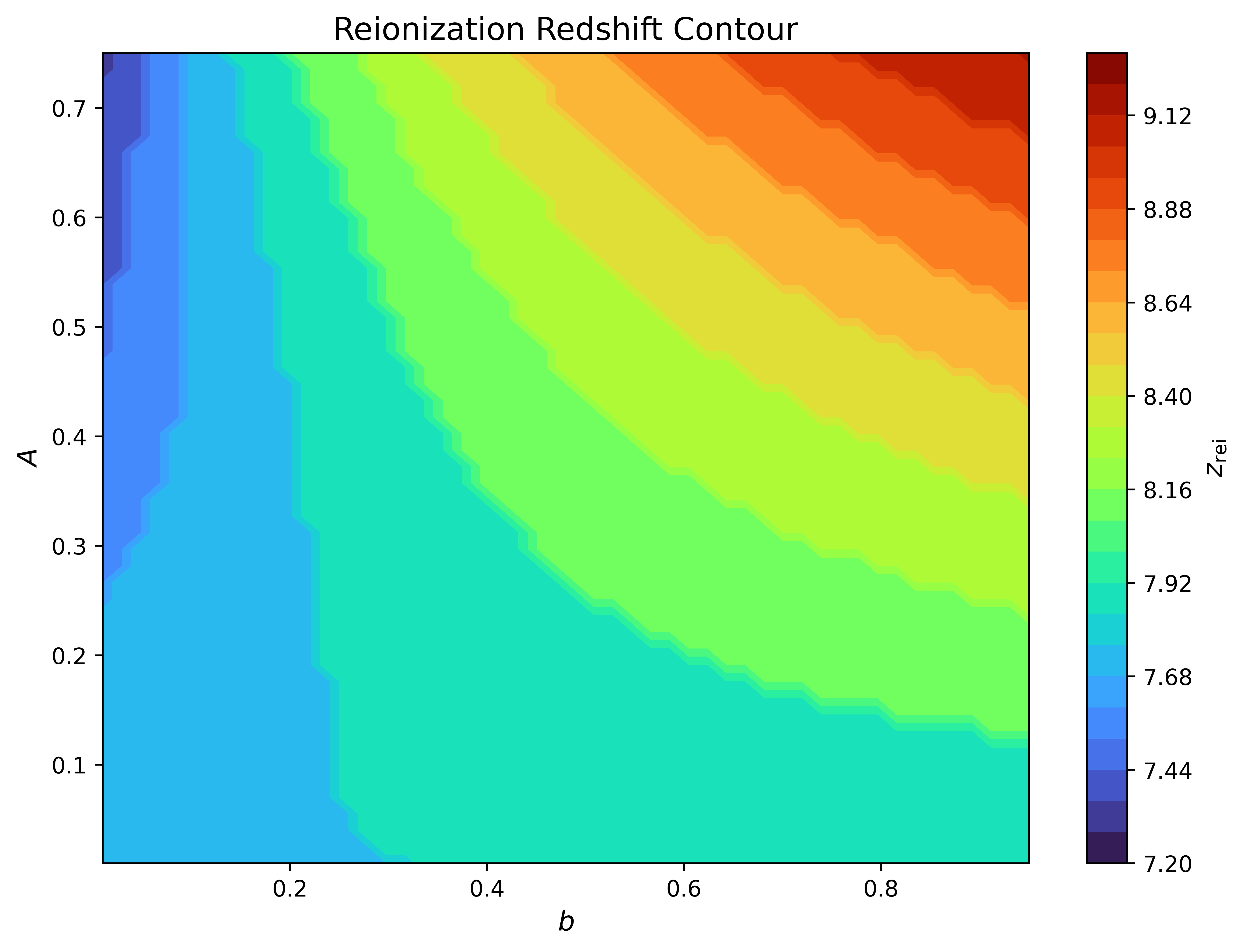}\\
	(a) Contours of the Thomson optical depth $\tau_{\rm{CMB}}$ & (b) Contours of the reionization redshift $z_{\rm rei}$\\ 
    in the $(A,b)$ parameter space. & in the $(A,b)$ parameter space.\\
    \end{tabular}
\caption{\label{fig:A_b_contours} Dependence of the reionization observables on the self-interaction gravity parameters $A$ and $b$, with all other model parameters fixed. The left panel shows the predicted Thomson optical depth, while the right panel shows the corresponding reionization completion redshift.}
\end{figure*}

\autoref{fig:A_b_contours}\added{ ,\ shows the sensitivity of the Thomson optical depth, $\tau_{\rm CMB}$, and the reionization completion redshift, $z_{\rm rei}$, to the depletion-function parameters $(A,b)$, with the transition parameters fixed at $z_g=4$ and $\tau=3$. At fixed $A$, the variation of $z_{\rm rei}$ with $b$ broadly follows the trend seen in the $\tau_{\rm CMB}$ contour, with both observables generally increasing as $b$ increases. This correspondence suggests that the two observables respond in a broadly consistent manner to variations in the late-time depletion parameters. However, the variation in $z_{\rm rei}$ across the $(A,b)$ plane is relatively modest compared with that across the $(zg,\tau)$ plane in \autoref{fig:zg_tau_contours}. This contrast indicates that the transition parameters have a stronger influence on the reionization completion epoch than the late-time parameters do.} 
\par \added{ The differing responses of the two observables $(z_{rei},\tau_{\rm CMB})$ in \autoref{fig:zg_tau_contours} may be understood in terms of the stronger influence of the transition parameters $(z_g,\tau)$ on the expansion history, which can modify the Hubble-rate weighting in the integrated optical depth. In contrast, the late-time parameters $(A,b)$ produce a closer correspondence between the variations in $z_{\rm rei}$ and $\tau_{\rm CMB}$ in \autoref{fig:A_b_contours}, suggesting a comparatively weaker influence of changes in the expansion history over the reionization epoch.}


\section{Observational Constraints}
\label{Sec:observational_constraints}

This section presents the observational constraints on the SI gravity model obtained \added{from the combined Union3 Type-Ia Supernova \cite{union3}, DESI DR2 BAO \cite{DESI_DR2}, CMB acoustic scale ($\theta_{*}$) data sets.} We first describe how the modified expansion history is translated into cosmological observables for each data set, then construct the likelihood function and the adopted parameter space and priors. We then present the Markov Chain Monte Carlo (MCMC) constraints on the depletion function parameters and assess the performance of the SI gravity model relative to the standard spatially flat $\Lambda$CDM cosmology using statistical model selection criteria.

The depletion function $D_{\rm M}(z)$ modifies the Hubble expansion rate through the modified Friedmann equation described in \autoref{eq:Ez_matter}. All cosmological distance measures entering the likelihood analysis are therefore computed using the corresponding modified Hubble parameter,
\begin{equation}
H(z)=H_0E(z),
\label{eq:Hubble_parameter}
\end{equation}
where $E(z)$ is evaluated within the SI gravity framework.

\subsection{Union3 Type-Ia Supernova Likelihood}
\label{subsec:union3_likelihood}

We employ the Union3 Type-Ia supernova compilation \cite{union3}, containing 2087 spectroscopically confirmed Type-Ia supernovae spanning the redshift range $0.01<z<2.26$.

Instead of working with the full unbinned sample, we use the publicly available compressed Union3 likelihood, derived from the complete Union3 analysis within the UNITY1.5 Bayesian framework, which provides 22 binned distance moduli and their covariance matrix \cite{union3}. This compressed likelihood essentially preserves all background-expansion cosmological information while greatly reducing the computational cost of repeated evaluations.

For a given cosmological model, the theoretical luminosity distance is computed as

\begin{equation}
d_L(z)
=
\frac{c(1+z)}{H_0}
\frac{1}{\sqrt{\Omega_k}}
\sinh\!\left[
\sqrt{\Omega_k}
\int_{0}^{z}
\frac{dz'}{E(z')}
\right],
\label{eq:luminosity_distance}
\end{equation}
where $E(z)$ denotes the normalized Hubble parameter and $\Omega_k$ is the curvature density parameter.

The corresponding theoretical distance modulus is

\begin{equation}
\mu_{\rm th}(z)
=
5\log_{10}
\left(
\frac{d_L}{10~{\rm pc}}
\right).
\label{eq:distance_modulus}
\end{equation}

Since the compressed Union3 likelihood already incorporates the supernova standardization and calibration procedures, the theoretical distance modulus can be compared directly with the binned observations. The corresponding likelihood is written as

\begin{equation}
\chi^2_{\rm Union3}
=
\Delta\mu^{\rm T}
C^{-1}_{\rm Union3}
\Delta\mu,
\label{eq:chi2_union3}
\end{equation}
where

\begin{equation}
\Delta\mu
=
\mu_{\rm th}
-
\mu_{\rm obs}
-
\Delta M_B,
\label{eq:delta_mu}
\end{equation}
and $C_{\rm Union3}$ is the covariance matrix supplied with the compressed Union3 likelihood, including both statistical and systematic uncertainties. The terms $\mu_{\rm th}$ and $\mu_{\rm obs}$ in the expression of $\Delta\mu$ represent the theoretical and observed values of distance modulus, respectively. The nuisance parameter $\Delta M_B$, associated with the absolute supernova magnitude, is marginalized over during the likelihood evaluation \cite{union3}.

\subsection{DESI DR2 BAO Likelihood}
\label{subsec:desi_likelihood}

For the DESI DR2 BAO observations \cite{DESI_DR2}, the modified expansion history is used to compute the transverse comoving distance,

\begin{equation}
\mathcal{D}_M(z)
=
\frac{c}{H_0}
\frac{1}{\sqrt{\Omega_k}}
\sinh\!\left[
\sqrt{\Omega_k}
\int_0^z
\frac{dz'}{E(z')}
\right],
\label{eq:DM}
\end{equation}
together with the Hubble distance,

\begin{equation}
D_H(z)
=
\frac{c}{H(z)}
=
\frac{c}{H_0E(z)}.
\label{eq:DH}
\end{equation}

Whenever isotropic BAO measurements are considered, the corresponding volume-averaged distance is,

\begin{equation}
D_V(z)
=
\left[
z\mathcal{D}_M^2(z)D_H(z)
\right]^{1/3}.
\label{eq:DV}
\end{equation}

The DESI DR2 compilation contains different BAO observables depending on the tracer population and effective redshift, with all distances expressed in units of the comoving sound horizon at the baryon drag epoch, $r_d$. In particular, the low-redshift Bright Galaxy Survey (BGS) sample provides measurements of $D_V/r_d$, whereas the remaining galaxy and Ly$\alpha$ samples constrain the anisotropic quantities $\mathcal{D}_M/r_d$ and $D_H/r_d$. Accordingly, the theoretical prediction vector is constructed using the appropriate distance combination before comparison with the observational data.

The corresponding BAO likelihood is \cite{DESI_DR2}, 

\begin{equation}
\chi^2_{\rm BAO}
=
\Delta y^{\rm T}
C^{-1}_{\rm BAO}
\Delta y,
\label{eq:chi2_bao}
\end{equation}
where

\begin{equation}
\Delta y
=
y_{\rm th}
-
y_{\rm obs},
\label{eq:delta_y}
\end{equation}
with

\begin{equation}
y=
\begin{cases}
D_V/r_d, & \text{for BGS},\\[0.2cm]
\left(\mathcal{D}_M/r_d,\;D_H/r_d\right), & \text{otherwise},
\end{cases}
\label{eq:bao_observables}
\end{equation}
where $y$ denotes the BAO observable, with $y_{\rm th}$ and $y_{\rm obs}$ representing its theoretical and observed values, respectively. The covariance matrix $C_{\rm BAO}$, provided by the DESI  \cite{DESI_DR2}, incorporates both statistical and systematic uncertainties, as well as the correlations between different tracers and redshift bins.

\subsection{\added{CMB acoustic scale constraint}}
\label{subsec:cmb_likelihood}
\added{The constraints obtained from Union3 Type-Ia Supernova \cite{union3} and DESI DR2 BAO \cite{DESI_DR2} are supplemented by a consistency test based on the angular acoustic scale of the cosmic microwave background (CMB) \cite{plank_pr3}. This provides an additional test of the expansion history up to the epoch of photon decoupling. The angular acoustic scale is defined as,}
\added{\begin{equation}
\theta_* = \frac{r_s(z_*)}{\mathcal{D}_M(z_*)},
\end{equation}}
\added{where $r_s(z_*)$ is the comoving sound horizon at photon decoupling and $\mathcal{D}_M(z_*)$ is the transverse comoving distance to that epoch. We adopt $z_*=1089$, corresponding approximately to the redshift of photon decoupling.}

\added{In the present analysis, we adopt $r_s(z_*)=144\,\mathrm{Mpc}$, motivated by the early Universe behavior discussed in a previous work \cite{Sargent2024}. In the SI gravity prescription considered here, the depletion correction becomes negligible at sufficiently high redshift, suggesting that its effect on the early-time acoustic evolution may be small. The adopted sound horizon is therefore assumed to be approximately equal to its standard cosmological value.} 

\added{The transverse comoving distance $\mathcal{D}_M(z_*)$ is calculated using the curvature-corrected distance relation introduced in Section \ref{subsec:desi_likelihood}, with the Hubble expansion rate specified by the SI gravity background and the corresponding spatial-curvature prescription. Combining this distance with the adopted sound horizon gives the model acoustic scale as,}
\added{\begin{equation}
\theta_*^{\rm model}
= \frac{144\,\mathrm{Mpc}}{\mathcal{D}_M(z_*)}.
\end{equation}}
\added{We compare this quantity with the observed CMB acoustic scale. For the Planck $PR3$ baseline analysis \cite{plank_pr3}, the measured value is $100\theta_*=1.04109$, corresponding to $\theta_*^{\rm obs}=0.0104109$.}

\added{Assuming a Gaussian constraint on the measured acoustic scale, its contribution to the likelihood is written as,}
\added{\begin{equation}
\chi^2_{\theta_*}
= \left(
\frac{\theta_*^{\rm model}-\theta_*^{\rm obs}}
{\sigma_{\theta_*}}
\right)^2,
\end{equation}}
\added{where, $\sigma_{\theta_*}$ is the observational uncertainty adopted from the CMB constraint \cite{plank_pr3}. This contribution can be combined with the supernova and BAO likelihoods to assess how the acoustic scale measurement restricts the allowed SI gravity parameters. }
\par \added{In the SI gravity framework no dark matter or cosmological constant is assumed; the reference model is Einstein--de Sitter with $\Omega_M = 1$, and the screened densities $\Omega^*_M(z) = \Omega_M D_M(z)$ are effective dynamical quantities rather than physical densities \cite{Deur2019,Deur2022}. The physical baryon density is specified independently of the reference total matter density. Accordingly, $\Omega_M h^2$ should not be compared with the $\Lambda$CDM value of $\omega_{\rm m}$. For the reionization calculation we fix the physical baryon density to $\omega_{\rm b}=0.0224$ in both cosmologies. Because $D_M\to1$ at early times, previous SI analyses treat the pre-recombination physics as standard~\cite{Deur2022,Sargent2024}, which is the assumption behind the adopted value of $r_s(z_*)$ above. We note that using the standard value of $\omega_b$ and the associated sound horizon presumes that pre-recombination physics in SI gravity is close enough to that of standard $\Lambda$CDM. The $\theta_*$ constraint is thus conditional on standard pre-recombination physics, and a fully self-consistent treatment of $r_s$ and $\omega_b$, together with the complete compressed CMB likelihood $(\theta_*, \omega_b, \omega_{bc})$, would require the early-universe and perturbation sector of SI gravity, which we leave to future work.}

\subsection{Parameter Estimation}
\label{subsec:parameter_estimation}

Assuming Gaussian observational uncertainties, the total likelihood is expressed as

\begin{equation}
\mathcal{L}
\propto
\exp\left(-\frac{\chi^2}{2}\right),
\label{eq:likelihood}
\end{equation}
where

\added{\begin{equation}
\chi^2
=
\chi^2_{\rm Union3}
+
\chi^2_{\rm BAO}
+
\chi^2_{\theta_*}.
\label{eq:chi2_total}
\end{equation}}

The likelihood is constructed from the total $\chi^2$ statistics,
 which quantifies the goodness of fit between the model
predictions and the observational data. The posterior distribution of the cosmological parameters is sampled using the Bayesian Markov Chain Monte Carlo (MCMC) framework implemented in the \texttt{Cobaya} package \cite{Torrado_2021}. The MCMC analysis is performed using the Metropolis sampler, while the resulting chains are analysed and visualised with the \texttt{GetDist} package~\cite{Lewis_2025} to obtain the marginalized posterior distributions, confidence intervals, and parameter correlation contours. Convergence of the MCMC chains is assessed using the Gelman--Rubin
convergence criterion, requiring the Gelman--Rubin statistic $R$ to satisfy
$|R-1|<0.01$ for all sampled parameters.

To provide a consistent assessment of the SI gravity framework, identical likelihood functions constructed from the \added{Union3, DESI DR2, and CMB acoustic scale($\theta_*$) data sets} are employed for both the SI gravity model and the standard spatially flat $\Lambda$CDM cosmology. For the baseline $\Lambda$CDM model, the sampled parameter vector is
\begin{equation}
\Theta_{\Lambda{\rm CDM}}
=
\left\{
\Omega_m,\,\added{H_0}\,,
H_0r_d
\right\},
\label{eq:lcdm_parameters}
\end{equation}
where $\Omega_m$ denotes the present-day matter density parameter, \added{ $H_0$ is the present-day Hubble parameter} and the combination $H_0r_d$ determines the overall BAO distance scale.

The SI gravity model extends the baseline cosmology by introducing the four depletion-function parameters, leading to the parameter vector

\begin{equation}
\Theta_{\rm SI}
=
\left\{
z_g,\,
\tau,\,
A,\,
b,\,\added{H_0,}\,
H_0r_d
\right\},
\label{eq:si_parameters}
\end{equation}
where $z_g$ specifies the characteristic redshift at which gravitational depletion becomes significant, $\tau$ controls the transition width, $A$ determines the depletion amplitude, and $b$ governs the steepness of the transition. These parameters completely characterize the phenomenological depletion function entering the modified Friedmann equation described in \autoref{Sec:si_gravity}. \added{The present-day Hubble parameter $H_0$ is included as a sampled parameter to characterize the expansion rate, while the combination $H_0 r_d$ is introduced as an auxiliary parameter required to evaluate the BAO distance observables and is not an intrinsic parameter of the SI gravity model itself.} Uniform priors are adopted for all sampled parameters over the intervals listed in Table~\ref{tab:priors}.

\begin{figure}
\includegraphics[width=0.95\textwidth]{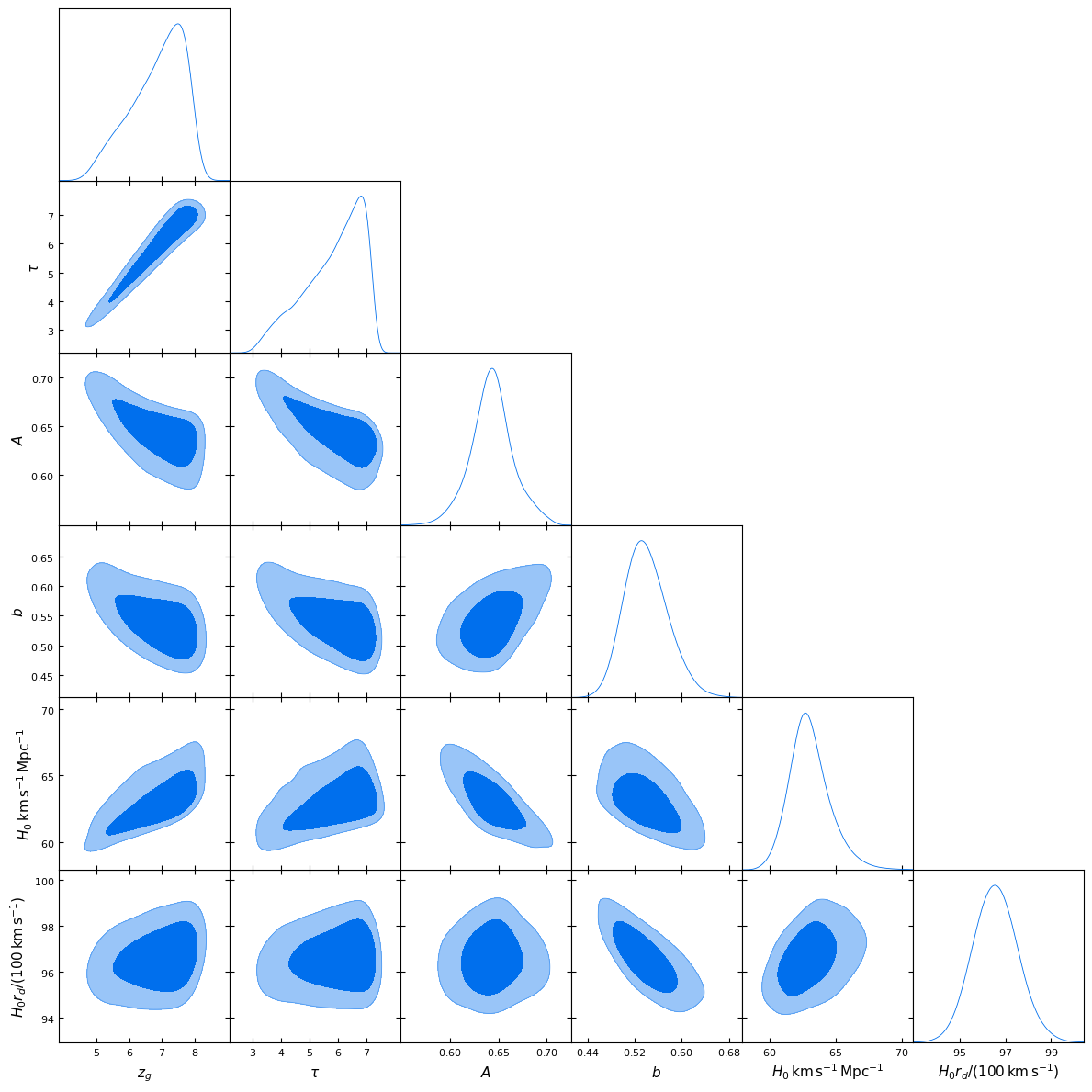}
\caption{Corner plot showing the marginalized posterior distributions and parameter correlations obtained from the MCMC analysis of the SI gravity model using the combined Union3 \cite{union3}, DESI DR2 \cite{DESI_DR2},\added{CMB acoustic scale($\theta_*$) \cite{plank_pr3} } datasets. The sampled parameters are the depletion function parameters $(z_g,\tau,A,b)$ together with \added{$H_0$ and} the dimensionless BAO scale parameter $H_0r_d/(100\,\mathrm{km\,s^{-1}})$. The diagonal panels show the one-dimensional marginalized posterior distributions, while the off-diagonal panels display the corresponding two-dimensional posterior distributions. The dark and light shaded regions denote the 68\% and 95\% credible intervals, respectively.
}
\label{fig:corner_plot}
\end{figure}

\begin{table*}[h!]
\centering
\caption{Uniform priors adopted for the MCMC analysis. \added{The parameters $H_0r_d$ and $H_0$ are included in the analysis, with $H_0r_d$ characterizing the overall BAO distance scale and $H_0$ denoting the Hubble constant.}}
\label{tab:priors}
\begin{tabular}{lc}
\hline\hline
Parameter & Uniform Prior \\
\hline
$z_g$ & $[1,\;50]$ \\
$\tau$ & $[1,\;50]$ \\
$A$ & $[0.01,\;1]$ \\
$b$ & $[0.01,\;1]$ \\
\added{$H_0\,(km\,s^{-1}\,Mpc^{-1})$} & \added{$[50,\;100]$} \\
$H_0r_d/\rm (100 km\,s^{-1}$) & $[80,\;120]$ \\
\hline\hline
\end{tabular}
\end{table*}

\autoref{fig:corner_plot} presents the marginalized posterior distributions obtained from the MCMC analysis. The diagonal panels display the one-dimensional marginalized posterior distributions for each parameter, while the off-diagonal panels show the corresponding two-dimensional joint confidence contours at the 68\% and 95\% confidence levels. The contours illustrate the correlations among the depletion-function parameters and their degeneracies with the BAO distance-scale parameter. A particularly notable feature is the pronounced anti-correlation between the late-time depletion parameter $b$ and $H_0r_d$, with larger values of $b$ corresponding to smaller values of $H_0r_d$. The MCMC analysis favours $H_0r_d = 96.6\times10^2~{\rm km\,s^{-1}}$ compared with the
$\Lambda$CDM value of $H_0r_d = 101.018\times10^2~{\rm km\,s^{-1}}$ obtained from the same \added{Union3, DESI DR2, and CMB acoustic scale($\theta_*$) dataset}. Within the SI gravity model, this shift is accompanied by a posterior preference for larger values of the depletion-function parameter $b$. As a result, the exponential correction term $Ae^{-z/b}$, which is introduced phenomenologically to model the partial restoration of symmetry at late times in structure formation, becomes increasingly important for determining the observationally favoured depletion function. The tendency toward larger preferred values of $b$ indicates that this correction stays relevant and thus provides a non-negligible contribution to the reconstructed expansion history supported by the combined Union3, DESI DR2, \added{ and the CMB acoustic scale($\theta_*$) constraints.}

\begin{figure*}
\includegraphics[width=0.75\textwidth]{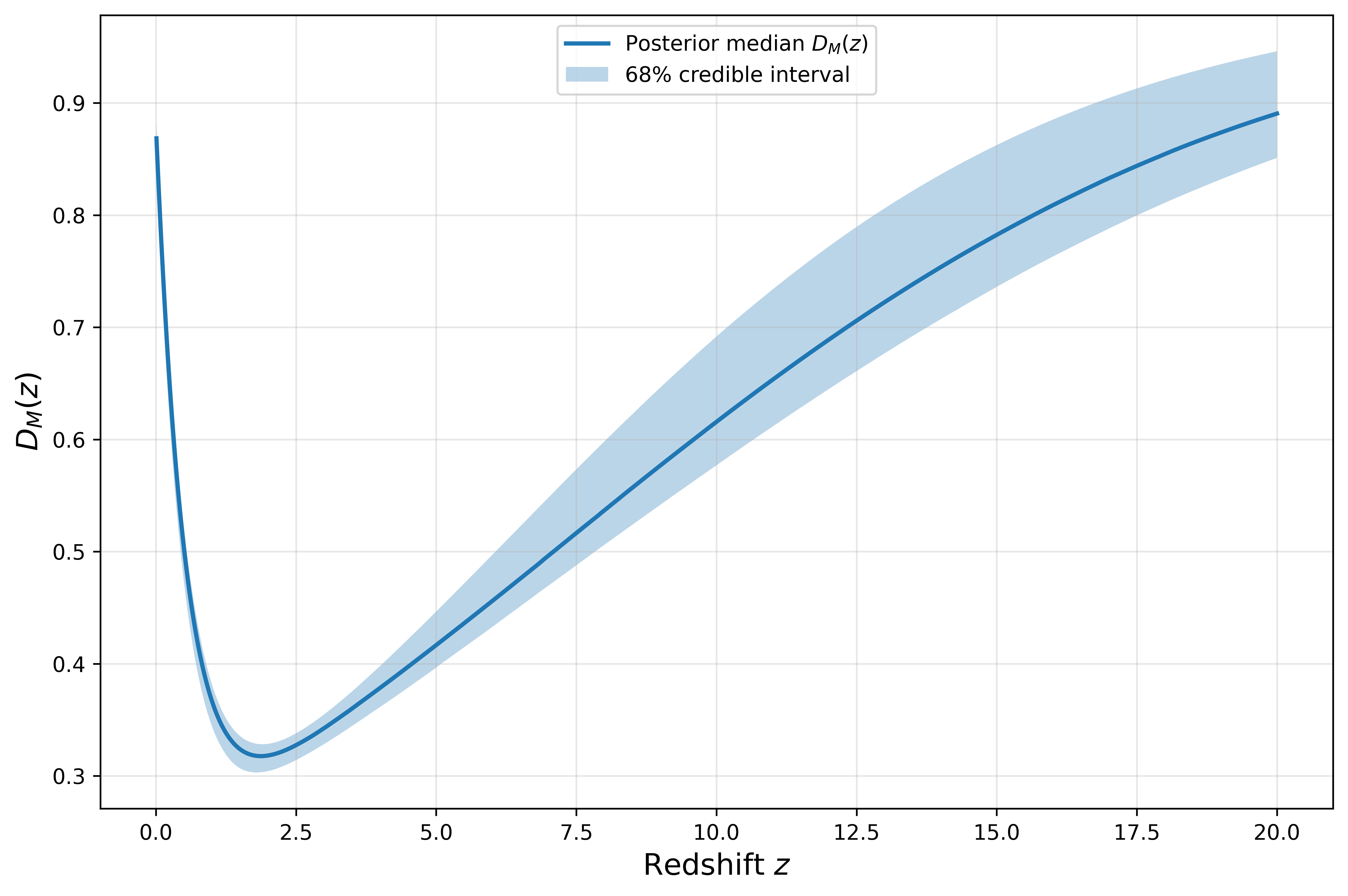}
\caption{Reconstructed depletion function $D_M(z)$ inferred from the joint Union3, DESI DR2, \added{and the CMB acoustic scale($\theta_*$)} analysis \added{of the adopted} SI gravity framework. The solid curve shows the \added{median depletion function}, while the shaded region denotes the corresponding 68\% credible interval obtained from the MCMC posterior samples. The reconstruction indicates that the depletion function approaches the homogeneous limit at high redshift and progressively decreases as nonlinear structures form, with the late-time exponential correction partially restoring $D_M(z)$ towards the present epoch.}\label{fig:depletion_function}
\end{figure*}

To illustrate the physical implications of these constraints, \autoref{fig:depletion_function} shows the reconstructed depletion function $D_M(z)$ obtained from the MCMC posterior. The solid curve corresponds to the \added{median}, while the shaded region represents the 68\% credible interval propagated from the posterior samples. As expected, $D_M(z)$ approaches unity at high redshift, reflecting the nearly homogeneous and isotropic Universe prior to significant structure formation. As nonlinear structures develop, gravitational field trapping progressively suppresses $D_M(z)$, while the late-time exponential correction partially restores the depletion function towards unity. The reconstructed evolution, therefore, provides a direct visualization of how the depletion function preferred by the combined Union3 \cite{union3}, DESI DR2 \cite{DESI_DR2}, \added{and the CMB acoustic scale($\theta_*$)} observations modifies the late-time expansion history within the SI gravity framework.

\begin{table*}[h!]
\centering
\caption{Marginalized parameter constraints obtained from the joint Union3, DESI DR2 \added{and the CMB acoustic Scale($\theta_*$)} analysis. Quoted uncertainties correspond to the $68\%$ confidence intervals.}
\label{tab:constraints}
\begin{tabular}{lcc}
\hline\hline
Parameter & Model constraint ($68\%$ C.L.) \\
\hline
$z_g$  & $6.9^{+1.1}_{-0.5}$ \\
$\tau$ & $5.9^{+1.3}_{-0.6}$ \\
$A$  & $0.643\pm0.022$\\
$b$  & $0.540^{+0.032}_{-0.041}$ \\
\added{$H_0\,(km\,s^{-1}\,Mpc^{-1})$} & \added{$63.1^{+1.4}_{-1.7}$} \\
$H_0r_d/\rm (100 \ km\,s^{-1})$  & $96.6\pm0.99$ \\
\hline\hline
\end{tabular}
\end{table*}

The corresponding parameters with their median values together with their marginalized $68\%$ confidence intervals are summarized in Table~\ref{tab:constraints}.

\subsection{Model Comparison}
\label{subsec:model_comparison}

To quantify the relative performance of the SI gravity model with respect to the standard spatially flat $\Lambda$CDM cosmology, we compare both the minimum chi-square statistic and the Deviance Information Criterion (DIC).

The difference in the minimum chi-square values is defined as

\begin{equation}
\Delta\chi^2
=
\chi^2_{\rm SI,min}
-
\chi^2_{\Lambda{\rm CDM,min}},
\label{eq:delta_chi2}
\end{equation}
where a negative value indicates that the corresponding model \added{provides a better fit to the combined Union3, DESI DR2 and CMB acoustic scale($\theta_*$) observations.}

Since the SI gravity model contains additional depletion-function parameters, an improvement in the minimum $\chi^2$ alone is insufficient to establish a statistically preferred model. We therefore employ the Deviance Information Criterion, which accounts for the trade-off between goodness-of-fit and model complexity.

The DIC is defined as \cite{Spiegelhalter2002,Liddle2007},
\begin{equation}\vspace{0.01cm}
{\rm DIC} = 2\overline{\chi^2} -\chi^2_{\rm min},
\label{eq:dic}
\end{equation}\vspace{0.1cm}
where $\overline{\chi^2}$ denotes the posterior mean of the chi-square statistic.
The relative model preference is quantified through,
\begin{equation}\vspace{0.1cm}
\Delta{\rm DIC}
={\rm DIC}_{\rm SI} - {\rm DIC}_{\Lambda{\rm CDM}}.
\label{eq:delta_dic}
\end{equation}\vspace{0.01cm}

\begin{table*}[h!]\vspace{0.01cm}
\centering
\caption{Model comparison between the SI gravity framework and the standard spatially flat $\Lambda$CDM cosmology.}
\label{tab:model_comparison}
\begin{tabular}{lcccc}
\hline\hline
Model &
$\chi^2_{\rm min}$ &
DIC &
$\Delta\chi^2$ &
$\Delta{\rm DIC}$ \\
\hline
$\Lambda$CDM & $38.83$ & $48.21$ & $0$ & $0$ \\
SI Gravity & $33.75$ & $48.13$ & $-5.08$ & $-0.08$ \\
\hline\hline
\end{tabular}
\end{table*}
\par Negative values of both $\Delta\chi^2$ and $\Delta{\rm DIC}$ indicate a statistical preference for the SI gravity model over the $\Lambda$CDM cosmology. While $\Delta\chi^2$ measures the improvement in fit quality, the DIC assesses whether this improvement remains significant after accounting for the additional model complexity. The resulting values of $\chi^2_{\rm min}$, DIC, $\Delta\chi^2$, and $\Delta{\rm DIC}$ for both cosmological models are summarized in Table ~\ref{tab:model_comparison}.

\added{For the present analysis, the SI gravity model yields $\Delta\chi^2\simeq-5.08$ and $\Delta\mathrm{DIC}\simeq-0.08$ when constrained by the combined Union3, DESI DR2, and CMB acoustic scale ($\theta_*$) data. The lower $\chi^2$ indicates that the SI gravity model provides a better fit to the combined observations than the baseline spatially flat $\Lambda$CDM cosmology. Although the nearly vanishing $\Delta\mathrm{DIC}$ does not establish a statistically meaningful preference between the two models after accounting for their effective complexity, the improved best fit suggests that the SI gravity motivated framework remains a viable alternative.}

\section{Implications for Cosmic reionization}
\label{Sec:jwst_cmb_reconsiln}

\added{The joint analysis of the Union3 \cite{union3}, DESI DR2 BAO measurements \cite{DESI_DR2}, and the CMB acoustic scale($\theta_*$) constraint \cite{plank_pr3} described in \autoref{Sec:observational_constraints}} determines the observationally preferred expansion history of  \added{the adopted} SI gravity framework. Having constrained the depletion-function parameters, we now investigate the implications of the resulting expansion history for cosmic reionization. Throughout this analysis, the astrophysical ingredients are kept fixed, with both the SI gravity and $\Lambda$CDM cosmologies employing the same updated JWST ultraviolet luminosity function, ionizing photon production efficiency, and escape fraction described in \autoref{Sec:reionization}. \added{The observed JWST ultraviolet luminosity function is, however, mapped consistently for SI gravity using the prescription described in \autoref{Sec:par_explrn}.}  \added{As a result, the variations in the inferred reionization history and Thomson optical depth stem from the altered expansion history in SI gravity and the related cosmology-dependent UVLF remapping, not from any adjustments to the assumed astrophysical prescriptions.}

\begin{figure*}
\includegraphics[width=0.8\linewidth]{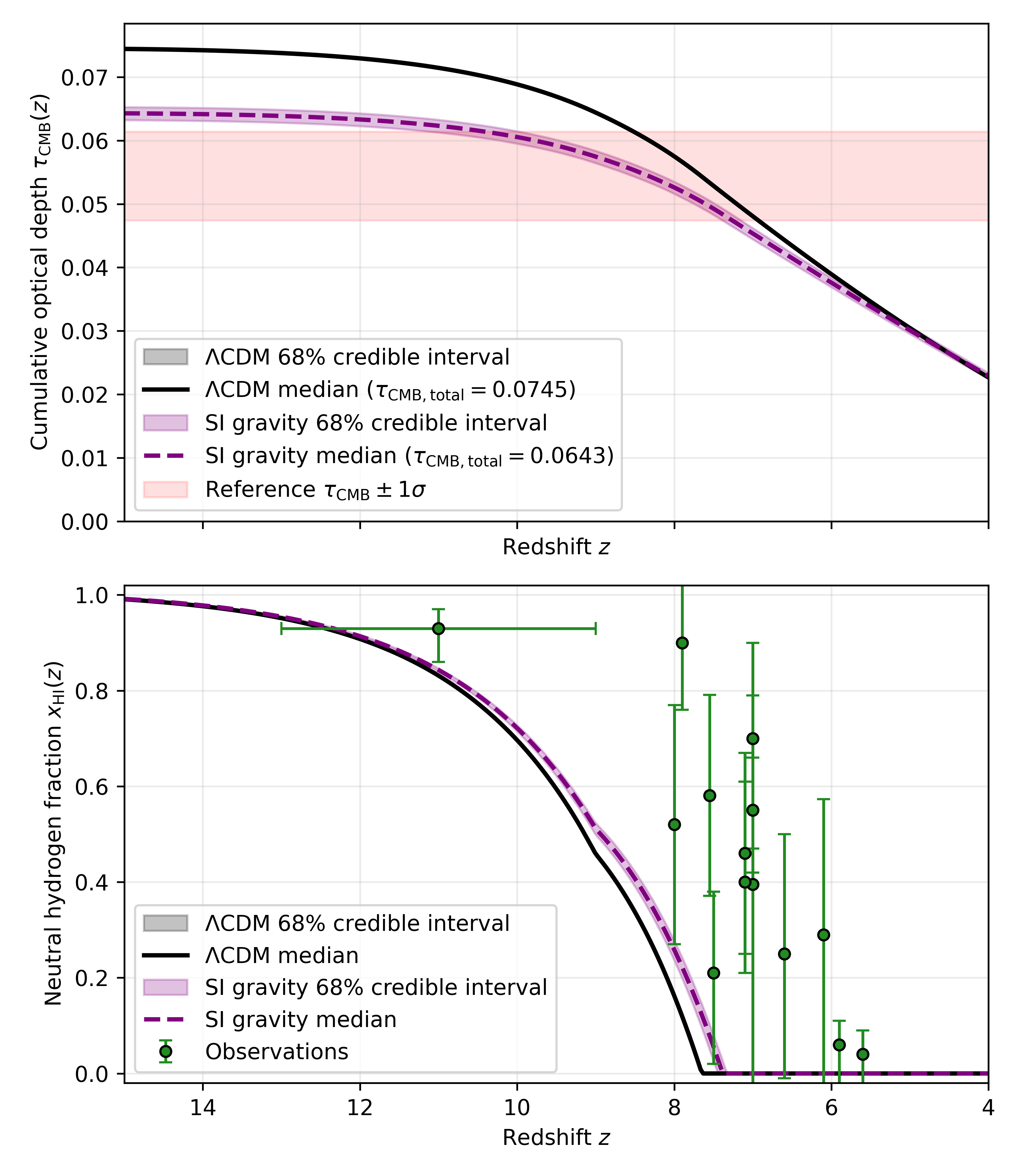}
\caption{Comparison of the reionization predictions for the SI gravity and $\Lambda$CDM cosmologies. The upper panel shows the Thomson optical depth together with the Planck PR3 $1\sigma$ constraint \citep{plank_pr3}. The lower panel shows the neutral hydrogen fraction, $x_{\rm HI}(z)$, compared with the observational compilation \cite{ly1,ly2,ly3,ly4,ly5,ly6,ly7,ly8}. Both cosmological models employ identical JWST-calibrated ultraviolet luminosity functions, ionizing photon production efficiencies, and escape fraction\added{, while mapping the JWST-calibrated ultraviolet luminosity function consistently for SI gravity.}}
\label{fig:reionization}
\end{figure*}

The resulting reionization histories are shown in \autoref{fig:reionization}, \added{with the cosmological parameters for both $\Lambda$CDM and SI gravity adopted from their respective MCMC analyses. For SI gravity, the adopted parameter values are summarized in Table~\ref{tab:constraints}. For $\Lambda$CDM model the model parameters are also adopted from the MCMC analysis yielding $\Omega_{m}=0.304\pm0.008$ and $H_0=68.83^{+0.77}_{-0.76}\,km\,s^{-1}\,Mpc^{-1}$.} As shown in the lower panel, the modified expansion history predicted by SI gravity produces only a slightly delayed reionization process compared with the standard $\Lambda$CDM cosmology. Due to this small shift, the predicted neutral-fraction evolution remains consistent with the current observational constraints from JWST over the observed redshift range and in tension with the observational neutral-fraction measurements compiled from quasar damping-wing, $\rm{Ly} \alpha$ emitter, and gamma-ray burst analyses \cite{ly1,ly2,ly3,ly4,ly5,ly6,ly7,ly8}.

The corresponding cumulative Thomson optical depth is shown in the upper panel of \autoref{fig:reionization}. Since the optical depth is determined by both the free-electron density and the Hubble expansion rate along the line of sight, the modified expansion history affects the line-of-sight integration at every redshift, yielding a measurable change in the final optical depth even though the ionization history differs only slightly from that of $\Lambda$CDM. \added{Under the adopted astrophysical prescriptions, the $\Lambda$CDM model yields a final Thomson optical depth of $\tau_{\rm CMB}=0.0745\pm0.0001$, whereas the SI-motivated model yields $\tau_{\rm CMB}=0.0643\pm0.001$. Relative to the Planck CMB constraint, this corresponds to a reduction in the discrepancy from approximately $3\sigma$ for $\Lambda$CDM model to $1.4\sigma$ for SI gravity model. These results indicate that the modified expansion history, together with the associated cosmology dependent UVLF remapping, can substantially alleviate the  CMB reionization optical depth tension.} This better agreement with the Planck PR3 measurement is achieved without modifying the astrophysical description of reionization, indicating that the change \added{can be traced to} the background expansion history, inferred from \added{the joint analysis of Union3 \cite{union3}, DESI DR2 \cite{DESI_DR2}, and the CMB acoustic scale data \cite{plank_pr3}, together with the associated cosmology-dependent UVLF remapping of the SI gravity model.}

\added{These findings show how cosmological bounds from low redshift data, combined with the CMB acoustic scale measurement, can be used to infer the expected evolution of cosmic reionization. The depletion-function parameters are constrained via a joint analysis of the Union3 \cite{union3} and DESI DR2 \cite{DESI_DR2} datasets, together with the CMB acoustic scale constraint \cite{plank_pr3}. The inferred expansion history is then employed to compute the reionization history and the integrated Thomson optical depth, incorporating the cosmology-dependent UVLF remapping and using identical astrophysical prescriptions for both cosmological models. Thus, the predicted optical depth is not obtained by directly fitting the high-redshift reionization observations examined here; instead, it emerges from the constrained cosmological background, the UVLF mapping method, and the adopted astrophysical assumptions.} Although the predicted neutral-fraction evolution remains in tension with the neutral-fraction measurements inferred from quasar damping-wing, $\mathrm{Ly}\,\alpha$ emitter, and gamma-ray burst observations \cite{ly1,ly2,ly3,ly4,ly5,ly6,ly7,ly8}, its consistency with the current JWST constraints, together with the agreement of the predicted optical depth with the Planck PR3 measurement \added{at the $1.4\sigma$ level}, indicates that the observationally constrained \added{Hubble evolution using the adopted SI gravity parameterization} \added{ can alleviate the CMB optical depth tension, although some discrepancies with independent high redshift probes remain.}

\section{Conclusions}
\label{Sec:conclusion}

\par In this work, we have investigated the emerging JWST–CMB discrepancies in the reionization history and the Thomson optical depth \added{in an expansion history motivated by self-interaction (SI) gravity.} Recent JWST observations suggest a substantially larger ionising-photon budget than previously inferred \cite{DONNAN2024}, owing to enhanced ultraviolet luminosity functions \cite{DONNAN2024} and ionising-photon production efficiencies, together with escape fractions motivated by observations of local analogues of high-redshift galaxies \cite{Chisholm2022}. Hence, these observations predict an earlier and more efficient reionization history, leading to a Thomson optical depth that exceeds the value inferred from CMB measurements in the $\Lambda$CDM framework \cite{Munoz2024}. Here we explore
whether this discrepancy \added{can be alleviated by modifying the cosmological expansion history, here through an SI gravity motivated depletion function, rather than the astrophysical description of reionization.} Within the SI gravity framework \cite{Deur2019}, leads to the trapping of gravitational fields inside nonlinear cosmic structures, thereby weakening gravity on cosmological scales and modifying the background expansion history.

\par The self-interaction (SI) gravity framework incorporates a depletion function into the Friedmann equation to model the confinement of gravitational fields and the resulting attenuation of the effective gravitational interaction on cosmological scales. In this work, we first study the \added{impact} of the depletion-function parameters $(z_g,\tau,A,b)$ on the Thomson scattering optical depth and on the redshift at which the intergalactic medium (IGM) becomes fully reionized. \added{Our results indicate that in the pre-MCMC exploration (Figures.~\ref{fig:zg_tau_contours}--\ref{fig:A_b_contours}), $\tau_{\rm CMB}$ varies mainly with the transition-duration parameter $\tau$ and only weakly with $z_g$, whereas $z_{\rm rei}$ responds to both $z_g$ and $\tau$ and varies over a wider relative range ($3.0\lesssim z_{\rm rei}\lesssim 9.75$ against $0.039\lesssim\tau_{\rm CMB}\lesssim 0.061$ in Figure.~\ref{fig:zg_tau_contours}). The parameters $(A,b)$ produces comparatively smaller variations in both quantities (Figure.~\ref{fig:A_b_contours}).} 

Subsequently, we constrain the SI gravity framework by performing a joint Bayesian analysis of the Union3 Type Ia supernova compilation \cite{union3}, DESI DR2 baryon acoustic oscillation (BAO) measurements \cite{DESI_DR2} \added{and the CMB acoustic scale($\theta_*$) measurement \cite{plank_pr3}}, thereby obtaining observational constraints on the depletion-function parameters. \added{Using the observationally constrained SI gravity model, we compute the ionization history and Thomson optical depth while keeping the UV luminosity functions, ionizing-photon production efficiencies, and escape fractions identical to those used in the $\Lambda$CDM case. The altered expansion history and the resulting cosmology-dependent UVLF remapping delay and modestly broaden reionization, lowering the integrated optical depth from $\tau_{\rm CMB}=0.0745\pm0.0001$ in $\Lambda$CDM to $\tau_{\rm CMB}=0.0643\pm0.001$ in SI gravity. This brings the prediction into ~ $(1.4\sigma)$ agreement with Planck PR3 \cite{plank_pr3}, versus a
~ $3\sigma$ $\Lambda$CDM mismatch, easing the optical-depth tension.}

\par \added{Moreover, the SI gravity motivated framework provides a better best-fit description of the combined cosmological observations than spatially flat $\Lambda$CDM, with $\Delta\chi^2\simeq-5.08$. However, the corresponding $\Delta\mathrm{DIC}\simeq-0.08$ indicates that this improvement is largely offset after accounting for the additional model complexity. Thus, while the SI gravity model achieves a better best fit, the information-criterion comparison does not establish a meaningful statistical preference over the standard $\Lambda$CDM model. These conclusions depend on the phenomenological depletion function, which is not derived from first principles for the SI background and perturbation equations; moreover, alternative modified expansion histories could influence the optical depth in a comparable manner. Nevertheless, the model remains a viable alternative framework for describing the cosmological observations considered here, warranting further investigation through independent observational tests and a more complete treatment of cosmological perturbations.} Overall, our results indicate that modifying only the cosmological background can substantially reduce the discrepancy between the optical depth inferred from JWST-inspired galaxy populations and that measured from the cosmic microwave background, without requiring any changes to the underlying astrophysical modelling of reionization. Nonetheless, the residual differences in the ionization history suggest that changes to the expansion history by themselves are unlikely to fully account for all reionization measurements, though further studies on various categories of models with altered Hubble evolution are required to unambiguously settle this issue.

\section{Acknowledgements}
\par \added{SM and SSP gratefully acknowledge the Council of Scientific and Industrial Research (CSIR), Government of India, for financial support through the CSIR-SRF-NET fellowship. The authors also thank the anonymous referee for valuable comments and suggestions that helped improve this manuscript.}

\bibliographystyle{JHEP}
\bibliography{refs.bib}

\end{document}